\documentclass[twocolumn,pra,aps,superscriptaddress,raggedbottom,showpacs,floatfix]{revtex4}

\usepackage{amsmath}
\usepackage{amssymb,amsthm}
\usepackage{latexsym}
\usepackage[dvips]{color}
\usepackage{graphicx}
\usepackage{bm}

\newcommand{\bra}[1]{\ensuremath{\langle{#1}|}}
\newcommand{\ket}[1]{\ensuremath{|{#1}\rangle}}
\newcommand{\opa}[1]{\ensuremath{{#1}}}
\newcommand{\mixr}{ {\scalebox{1}[1]{\ensuremath{\neg}}} }
\newcommand{\mixl}{ {\scalebox{-1}[1]{\ensuremath{\neg}}} }

\newcommand{\torder}[1]{\ensuremath{\opa{\text{T}}\left\{{#1}\right\}}}
\newcommand{\tcorder}[1]{\ensuremath{\opa{\text{T}_C}\left\{{#1}\right\}}}

\newcommand{\copa}[1]{\ensuremath{\hat{#1}}}
\newcommand{\copas}[1]{\ensuremath{#1}}

\newcommand{\acomm}[1]{\ensuremath{\left\{{#1}\right\}}}
\newcommand{\trace}[1]{\ensuremath{\text{tr}\left({#1}\right)}}
\newcommand{\traceb}[1]{\ensuremath{\text{tr}_\text{rest}\left({#1}\right)}}
\newcommand{\traceo}[1]{\ensuremath{\text{tr}_\text{0}\left({#1}\right)}}
\newcommand{\lwf}[1]{\ensuremath{\langle #1 \vert}}
\newcommand{\rwf}[1]{\ensuremath{\vert #1 \rangle}}
\newcommand{\cds}[1]{\ensuremath{\copas{c}^\dagger_{#1}}}
\newcommand{\ccs}[1]{\ensuremath{\copas{c}_{#1}}}
\newcommand{\cd}[1]{\ensuremath{\copa{c}^\dagger_{#1}}}
\newcommand{\cc}[1]{\ensuremath{\copa{c}_{#1}}}
\newcommand{\ads}[1]{\ensuremath{a^\dagger_{#1}}}
\newcommand{\aas}[1]{\ensuremath{a_{#1}}}

\newcommand{\nup}{\ensuremath{n_{\uparrow}}}
\newcommand{\ndown}{\ensuremath{n_{\downarrow}}}

\newcommand{\dt}{\ensuremath{\text{d}t}}

\newcommand{\exP}[1]{\ensuremath{\text{exp}\left({#1}\right)}}
\newcommand{\Exp}[1]{\ensuremath{\text{exp}\left[{#1}\right]}}

\newcommand{\abs}[1]{\ensuremath{\left|{#1}\right|}}

\newcommand{\est}[1]{\ensuremath{\langle {#1} \rangle}}

\newcommand{\mat}[1]{\ensuremath{\bm{{#1}}}}

\newcommand{\todo}[1]{}

\newcommand{\eqeqref}[1]{Eq.~\eqref{#1}}

\begin{document}

\title{Hamiltonian-based impurity solver for nonequilibrium dynamical mean-field theory}

\author{Christian Gramsch}
\affiliation{Theoretical Physics III, Center for Electronic Correlations and Magnetism, 
Institute of Physics, University of Augsburg, 86135 Augsburg, Germany}
\author{Karsten Balzer}
\affiliation{Max Planck Research Department for Structural Dynamics, University of Hamburg-CFEL,
22607 Hamburg, Germany}
\author{Martin Eckstein}
\affiliation{Max Planck Research Department for Structural Dynamics, University of Hamburg-CFEL,
22607 Hamburg, Germany}
\author{Marcus Kollar}
\affiliation{Theoretical Physics III, Center for Electronic Correlations and Magnetism, 
Institute of Physics, University of Augsburg, 86135 Augsburg, Germany}

\date{June 26, 2013}

\begin{abstract}
  We derive an exact mapping from the action of nonequilibrium
  dynamical mean-field theory (DMFT) to a single-impurity Anderson
  model (SIAM) with time-dependent parameters, which can be solved
  numerically by exact diagonalization. The representability of the
  nonequilibrium DMFT action by a SIAM is established as a rather
	general property of nonequilibrium Green functions.
	We also obtain the nonequilibrium DMFT equations using the
  cavity method alone. We show how to numerically obtain the SIAM
  parameters using Cholesky or eigenvector matrix decompositions. As
  an application, we use a Krylov-based time propagation method
  to investigate the Hubbard model in which the hopping is
  switched on, starting from the atomic limit. Possible future
  developments are discussed.
\end{abstract}


\maketitle

\section{Introduction}
Experiments on strongly correlated quantum many-body systems out of equilibrium have reached a high level 
of precision and control. One can excite 
complex  materials 
with femtosecond laser pulses and record
their subsequent time evolution on the timescale of the electronic motion~\cite{Cavalieri2007, Wall11}. In systems 
of ultra-cold atoms in optical lattices, on the other hand, interaction and bandwidth can be controlled as a 
function of time via Feshbach resonances and the depth of the lattice potential, respectively, and external 
fields can be mimicked by shaking or tilting the optical lattice~\cite{Bloch2008a,Struck2011,Simon2011}. 
The understanding of relaxation pathways in correlated systems touches upon fundamental questions of 
statistical mechanics~\cite{Polkovnikov2011RMP}, it can provide insights into the nature of correlated states 
which is not possible with conventional frequency-domain techniques, and it may lead to the discovery 
of `hidden phases', i.e., long-lived transient states that are inaccessible via any thermal pathway 
\cite{Fausti11,Ichikawa2011}. 

Stimulated by these developments, a growing theoretical effort is aimed at advancing the microscopic 
description of correlated lattice models out of equilibrium. A method which is well-suited to capture strong 
local correlation effects in higher-dimensional systems is the nonequilibrium formulation~\cite{Schmidt2002, 
Freericks2006} of dynamical mean-field theory (DMFT)~\cite{Georges96}. Over the past few years, 
nonequilibrium DMFT has been used in a large number of theoretical studies, including interaction quenches 
\cite{Eckstein09,Eckstein2010}, dc-field driven systems~\cite{Freericks2006,Freericks2008,Eckstein2011bloch,
Aron11,Amaricci2012}, photo-excitation of Mott insulators~\cite{Eckstein11,Moritz2010,Moritz2012,Eckstein2013} 
and nonequilibrium phase transitions from antiferromagnetic to paramagnetic states~\cite{Werner2012afm,Tsuji2013prl}. 

Within DMFT, a lattice model such as the Hubbard model is mapped onto an effective impurity model, which 
consists of a single site of the lattice coupled to a 
``noninteracting medium'' 
with which it can exchange 
particles. A big challenge for the advance of nonequilibrium DMFT is the development of appropriate 
methods for the solution of this single-impurity problem out of equilibrium. Continuous-time quantum 
Monte Carlo (CTQMC) on the Keldysh contour~\cite{Muehlbacher2008,Werner2009} can 
provide numerically exact DMFT results for short times~\cite{Eckstein09}, but 
the effort increases exponentially with time due to the phase problem. 
Several DMFT studies have instead used second- and third-order perturbation theory 
\cite{Eckstein2011bloch,Aron11,Amaricci2012,Tsuji2013prl}, which works in the weakly 
interacting regime, but can give unphysical results for larger values of the interaction due to its 
nonconserving nature~\cite{Eckstein2010,Tsuji2013}. Strong-coupling perturbation theory~\cite{Eckstein2010nca}, 
on the other hand, is suitable for the Mott insulator, but cannot address correlated metallic states
at low temperatures. Finally, a numerically tractable impurity model is obtained for the 
Falicov-Kimball model, where a solution is possible via a closed set of equations of motion 
\cite{Freericks2006,Freericks2008}. However, because only particles with one spin flavor can hop 
on the lattice in the Falicov-Kimball model, its dynamics are rather peculiar, which is reflected in the 
absence of thermalization of single-particle quantities~\cite{Eckstein2008a}. 

Impurity models are of interest 
also in their own right, apart from their importance for DMFT, e.g., for the description of quantum dots 
or Kondo impurities. Recently, sophisticated techniques such as diagrammatic (``bold-line'') CTQMC 
\cite{Gull2011bold} and influence functional approaches~\cite{Segal2010,Cohen2013} have been 
developed to study nonequilibrium dynamics motivated by transport experiments through quantum 
dots. While these techniques can address 
longer
times than direct CTQMC
simulations, they have not yet been used within DMFT, partly because the measurement of 
time-nonlocal correlation functions is technically challenging. Hence there is a clear need for the
development of novel impurity solvers for DMFT.

In equilibrium DMFT, a whole class of impurity solvers are based on a mapping of the impurity model to a suitable 
Hamiltonian representation: The effective medium of DMFT is approximated by a finite number of 
bath orbitals, and the resulting single-impurity Anderson model (SIAM) is solved using exact 
diagonalization~\cite{Georges96}, numerical renormalization group (NRG)~\cite{Bulla1999}, 
density-matrix renormalization group (DMRG)~\cite{Garcia2004}, or (restricted active-space) configuration-interaction 
approaches~\cite{Zgid2012}. 
Mapping
the DMFT impurity problem to a finite SIAM seems 
rather attractive 
for nonequilibrium studies, 
since it is not a priori restricted to either interaction or
hopping being small, and it can thus work also at intermediate coupling. 
Moreover,
it may even serve as a starting point for a diagrammatic expansion using the 
dual-fermion approach~\cite{Jung2012}.
However, the mapping procedure itself turns out to be more difficult for nonequilibrium situations than it 
is for equilibrium. In equilibrium, the mapping of the effective medium to a set of bath orbitals can be
performed by fitting the spectral function of the 
bath~\cite{Georges96}, while in nonequilibrium, 
when time-translational 
invariance is lost, a single spectral function is not enough to characterize a system. 
The situation is somewhat simpler in a steady state, which can be characterized by a spectral function and 
an occupation function (which is then different from the Fermi function). A representation of the impurity 
problem using a SIAM with dissipative (Markovian) terms has been discussed for this case~\cite{Arrigoni2013}.

In the present paper, we address the mapping problem for general time-evolving states. We prove that the 
nonequilibrium DMFT action for the Hubbard model in the limit of infinite dimensions can be represented by 
a SIAM, and we discuss methods to construct such a representation either exactly or approximately. In particular, 
we clarify how to separate the initial state correlations (which are represented by bath orbitals that are 
coupled to the impurity already before any 
perturbation), and  additional correlations 
that are built up 
at later times. First numerical tests for a quench in the Hubbard model from zero to finite hopping 
between the sites 
allow 
us to address a parameter regime of interactions which is not accessible 
with the presently available weak and strong-coupling solvers.

The outline of the paper is as follows: In Sec.~\ref{sec:mapping}, we give a brief overview on the DMFT equations
and the notation used for nonequilibrium Green functions. We then define the mapping problem
(Sec.~\ref{sec:mappingmapping}), and discuss the representability of the nonequilibrium DMFT action by a SIAM on 
general grounds (Sec.~\ref{sec:mappingrepresentation}). A derivation of nonequilibrium DMFT using the cavity method for 
contour Green functions is given in 
Sec.~\ref{sec:dmftaction}, with further details in Appendices~\ref{sec:genf} and \ref{ap:selfcondarblatt}.
In Sec.~\ref{sec:bathfitting}, we explain how to construct 
a Hamiltonian bath representation of the DMFT action. In Sec.~\ref{sec:approxrepr} we show how to do this in 
 practice, and we
present numerical results for the Hubbard model in nonequilibrium in Sec.~\ref{sec:bethequench}. 
Sec.~\ref{sec:summary} contains a summary.

\section{Nonequilibrium DMFT and the mapping problem}
\label{sec:mapping}

\subsection{Nonequilibrium Green functions}

Nonequilibrium DMFT is based upon the Keldysh formalism~\cite{Keldysh1964} for contour-ordered Green functions.
In the following subsection we briefly overview the basic concepts of the approach and define the relevant quantities 
for 
our 
subsequent analysis. A comprehensive introduction to the nonequilibrium many-body formalism itself can be 
found in a number of textbooks and review articles, e.g., the book by Kamenev for a general introduction 
\cite{KamenevBook}, and Ref.~\onlinecite{LeeuwenLectureNotes} for a detailed description of the formalism based onto 
the L-shaped time contour which will be used below.

\begin{figure}[!t]%
\includegraphics[width=0.97\columnwidth]{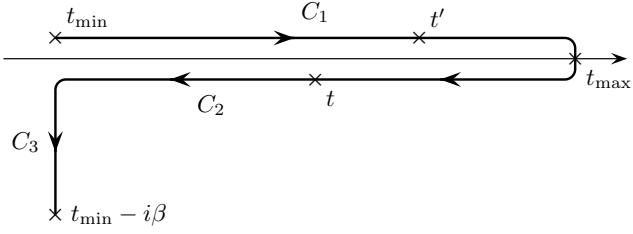}
\caption{The L-shaped integration contour $C$ consists of three parts. The integration runs along $C_1$ 
from $t_\text{min}$ to $t_\text{max}$ on 
the real axis. It then goes back along $C_2$, and follows $C_3$ into the complex plane 
to $t_\text{min}-i\beta$, where $\beta$ is the inverse temperature. 
In the sense of the contour,
the time $t\in C_2$ is therefore larger than $t'\in C_1$ in this example, i.e., $t>_Ct'$. 
For simplicity we 
set 
$t_\text{min}=0$ throughout this work.}%
\label{fig:Lshape}%
\end{figure}

From a general perspective, we would like to describe the real-time evolution of a quantum many-body system which 
is initially in thermal equilibrium at temperature $T=1/\beta$ (the initial state density matrix is $e^{-\beta H(0)}/Z$) and 
evolves unitarily under a time-dependent Hamiltonian $H(t)$ for times $t>0$. In specific applications we consider
the single-band Hubbard model 
\begin{align}
\label{eq:hubbard}
H_{\text{Hub}}(t) = \sum_{ij\sigma} t_{ij}(t)\, c_{i\sigma}^\dagger c_{j\sigma}
+ 
U(t)
\sum_{i}
(n_{i\uparrow}-\tfrac12)
(n_{i\downarrow}-\tfrac12),
\end{align}
where $c_{i\sigma}$ and $c_{i\sigma}^\dagger$ are annihilation and creation operators for an electron with spin 
$\sigma$ on site $i$ of a crystal lattice, $t_{ij}$ is the hopping matrix element between sites $i$ and $j$, and
$U$ is the local Coulomb repulsion. Within the nonequilibrium Green functions 
technique this problem is solved using time-dependent correlation functions with time arguments on the L-shaped 
time contour $C$ that runs from $0$ to $t_\text{max}$ (the maximum time of interest) along the real axis, 
back to time $0$, and to $-i\beta$ along the imaginary axis (Fig.~\ref{fig:Lshape}). 
The expectation value $\est{A(t)} = \text{tr}[e^{-\beta H(0)} U(t,0)^\dagger A U(t,0)]/Z$ of an observable $A$,
with the  unitary time-evolution operator $U(t,0) = T_t e^{-i\int_0^t  ds H(s)}$, can then be recast into the form 
of a contour-ordered expression  $\est{A(t)}_{S}\equiv\trace{\tcorder{\exp(S) A(t)}}/Z_{S}$, where the action is 
given by $S=-i\int_C\dt H(t)$, $\int_C\dt$ denotes the integral along the contour, $Z_{S}=\trace{\tcorder{\exp(S)}}$ 
is the partition function, and  $\tcorder{...}$ is the contour-ordering operator, which orders
operators according to their relative position on $C$,
\begin{equation}
	\tcorder{A(t)B(t')}=\left\{
	\begin{aligned}
		&\,\,\,AB &\text{ if }t>_Ct',\\
		&(\pm)BA &\text{ if }t<_Ct'.
	\end{aligned}
	\right.
	\label{eq:deftimedep}
\end{equation}
The negative sign applies when both $A$ and $B$ contain an odd number of fermionic 
annihilation or creation operators.

Green functions for a more general action $S$ are defined as 
similar
contour-ordered expectation values,
\begin{align}
\label{gdef}
	G_{\alpha\alpha'}(t,t')&=-i\est{\ccs{\alpha}(t)\cds{\alpha'}(t')}_{S}\\
	&\equiv\frac{-i}{Z_{S}}\trace{\tcorder{\exp(S)\ccs{\alpha}(t)\cds{\alpha'}(t')}}
	\nonumber,
\end{align}
where subscripts $\alpha$ and $\alpha'$ label spin and orbital quantum numbers. Depending on the time arguments, the entries of the contour-ordered Green function in Eq.~(\ref{gdef}) have a different physical 
meaning. Letting a superscript $a,b=1,2,3$ of $G^{ab}(t,t')$ indicate whether a time argument is on the upper ($1$), lower ($2$),
or imaginary ($3$) part of the contour, we define the 
following components (the subscripts for 
spin and orbital degrees 
of freedom are omitted for simplicity),
\begin{subequations}
\label{gcomponent}
\begin{align}
\label{gles def}
  G^<(t,t')
    &=
      G^{12}(t,t')
    =
     i \langle c^\dagger(t') c(t) \rangle_S,
  \\
\label{ggtr def}
  G^>(t,t')
    &=
      G^{21}(t,t')
    =
      -i \langle c(t) c^\dagger(t') \rangle_S,
\\
\label{eq:mixr}
G^ {\mixr}(t,\tau )
  &=
     G^{13}(t,\tau)
  =
    i \langle c^\dagger(\tau ) c(t) \rangle_S,
\\
\label{eq:mixl}
G^ {\mixl}(\tau, t)
  &=
     G^{31}(\tau,t)
  =
    -i\langle c(\tau) c^\dagger(t') \rangle_S,
\\
\label{gdef mat}
G^M(\tau)
  &=
    -iG^{33}(\tau,0)
  =
    -\langle c(\tau) c^\dagger(0) \rangle_S.
\end{align}
\end{subequations}
The first two functions, `lesser' and `greater', 
are related to photoemission and inverse photoemission, respectively~\cite{FreericksKrishnamurthyPruschke2009}. 
The mixed functions $G^ {\mixr}$ and $G^ {\mixl}$ encode correlations between the time-evolving state and the initial 
thermal equilibrium state, while the $33$-component is the Matsubara Green function of the initial thermal 
equilibrium state (in the definition of $G^M$, we have taken into account time-translational invariance of this
function).

The five components (\ref{gcomponent}) fully parametrize the Green function (\ref{gdef}), 
and all other entries can be restored by noting that the largest real-time argument can be shifted 
from the upper to the lower contour (because the time evolution on the forward and backward branch 
cancels). Other commonly used parameterizations involve the retarded and advanced functions,
$G^\text{ret}(t,t')=\Theta(t-t')[G^>(t,t')-G^<(t,t')]$ and  $G^\text{adv}(t,t')=\Theta(t'-t)[G^<(t,t')-G^>(t,t')]$, where $\Theta(t)$ denotes the Heaviside step function, 
but for this work we will only use the components defined in~(\ref{gcomponent}). Finally, we note 
the hermitian symmetry relation for the components,
\begin{subequations}
\begin{align}
  G^{<,>}(t,t')^\ast
    &=
      -G^{<,>}(t',t),
  \\
  G^ {\mixr}(t,\tau )^\ast
    &=
     G^ {\mixl}(\beta-\tau ,t),
  \label{h.c. G13}
\end{align}
\label{chermitian}
\end{subequations}
which will be important below.

For illustration and later reference, we give an analytic expression for the Green function of an isolated bath-orbital
with time-dependent energy $\epsilon(t)$,
\begin{align}
\label{gsignglebath1}
g(t,t')
=
-i
\frac{
\text{tr}\left(
\tcorder{e^{-i\int_C ds (\epsilon(s)-\mu) c^\dagger c}  c(t)c^\dagger(t')} 
\right)
}{
\text{tr}\left(
\tcorder{e^{-i\int_C ds (\epsilon(s)-\mu)c^\dagger c}}
\right)
}.
\end{align}
Using Heisenberg equations of motion for the operators $c(t)$, one can see that 
\begin{equation}
	\label{eq:gsinglesite}
	g(t,t')= -i\Big[\Theta_C(t,t')-f\big(\epsilon(0)-\mu\big)\Big]
	\mathcal{E}(t)\mathcal{E}(t')^{-1},
\end{equation}
where $f(\epsilon)$ $=$ $1/(e^{\beta\epsilon}+1)$ is the Fermi function,
\begin{equation}
	\mathcal{E}(t)=\left\{\begin{aligned}
		&\text{e}^{-i \int_0^t ds 
		\left[
		\epsilon(s)-\mu
		\right]
		} 
		\text{ ~~~for real }t\\
		&\text{e}^{-\tau 
		\left[
		\epsilon(0)-\mu
		\right]
		}
		\text{ ~~~for }t=-i\tau\\
	\end{aligned}\right.,
\end{equation}
and $\Theta_C(t,t')$ is the contour analog of the Heaviside step function, i.e.,
\begin{equation}
	\label{eq:Theta}
	\Theta_C(t,t')=\left\{\begin{aligned}
		&1\text{ ~~for }t\ge_Ct'\\
		&0\text{ ~~else. }
	\end{aligned}\right..
\end{equation}
The Green function for a single site with time-independent orbital energy $\epsilon(t)\equiv \epsilon$ 
will be denoted by
\begin{equation}
\label{g eps const}
	g(\epsilon,t,t') \equiv -i\left[\Theta_C(t,t')-f(\epsilon)\right]\text{e}^{-i \epsilon(t-t')}.
\end{equation}

\subsection{Nonequilibrium DMFT}

In DMFT, the Hubbard Hamiltonian (\ref{eq:hubbard}) is mapped onto an effective impurity model from which all local 
correlation functions~\cite{Georges96} can be obtained. The key step in DMFT is to compute the local Green function
\begin{equation}
\label{dmft g}
G_{\sigma}(t,t') = -i\est{\ccs{\sigma}(t)\cds{\sigma}(t')}_{S_\text{loc}}
\end{equation}
from a single-site model which is defined by the action
\begin{multline}
\label{eq:dmftaction}
S_\text{loc}=-i\int\limits_C\dt\biggl[U(t)(\nup(t)-\tfrac12)(\ndown(t)-\tfrac12)-\mu\sum_\sigma n_\sigma(t)\biggr]
\\
-i\int\limits_C
\int\limits_C
\dt_1\dt_2\sum_{\sigma}\Lambda_{\sigma}(t_1,t_2)\cds{\sigma}(t_1)\ccs{\sigma}(t_2).
\end{multline}
Here the first part contains the Hamiltonian of an isolated site of the original Hubbard Hamiltonian, while $\Lambda$ 
is the hybridization of that site with a noninteracting environment, which must be determined self-consistently. In 
Section~\ref{sec:bethequench}, we study the Bethe lattice with  nearest-neighbor hopping $t_{ij}$ $=$ $v/\sqrt{\mathcal{Z}}$ in the limit of 
infinite coordination number $\mathcal{Z}$~\cite{Metzner1989}, corresponding to a semielliptical density of states,
\begin{align}
\label{eq:bethed|os}
\rho(\epsilon) = \frac{\sqrt{4v^2-\epsilon^2}}{2\pi v^2}.
\end{align}
For time-dependent $v$, the  self-consistency can then be written in closed form~\cite{eck:09.09,epjst2010a,Eckstein2010ramps},
\begin{equation}
\label{bethe}
\Lambda_\sigma(t,t') =  v(t) G_\sigma(t,t') v(t').
\end{equation}
For a general lattice in the limit of infinite dimensions~\cite{Metzner1989}, where DMFT becomes exact, a formal 
expression for the hybridization function at a given site $0$ can be given in terms of the cavity Green function 
$G^{(0)}$, i.e., the Green function of the original Hubbard Hamiltonian from which site $0$ has been removed,
\begin{align}
	\label{eq:cavity}
	&\Lambda_{\sigma}(t,t')=\sum_{i,j}t_{0i}(t) G^{(0)}_{ij\sigma}(t,t')t_{j0}(t').
\end{align}
The derivation of this expression for the effective nonequilibrium DMFT action via the cavity method largely 
parallels the corresponding formulation for Matsubara Green functions~\cite{Georges96} and is presented 
separately in Sec.~\ref{sec:dmftaction}.

In general, the evaluation of the DMFT self-consistency depends on the lattice structure and on external fields, 
and \eqeqref{eq:cavity} cannot be recast in closed form like \eqeqref{bethe}. However, the precise form of the implicit functional relation 
$\Lambda=\Lambda[G]$ is not important for the exact-diagonalization-based impurity solver developed in this paper, 
and we thus refer to the literature \cite{Freericks2006,Eckstein11} and Appendix~\ref{ap:selfcondarblatt}
for detailed descriptions of the nonequilibrium DMFT equations.

\subsection{The mapping problem}
\label{sec:mappingmapping}

Because of the interaction $U(t)$, it is a complicated problem to calculate $G_\sigma(t,t')$ for a given 
$\Lambda_\sigma(t,t')$ from the action~\eqref{eq:dmftaction}. For the corresponding equilibrium action a 
mapping to an appropriate single-impurity Anderson model (SIAM) is very successful~\cite{Georges96}, 
because this allows the use of Hamiltonian-based solvers for   
the calculation of the local Green function. The 'mapping problem' which we address 
in the following
refers to a similar construction for nonequilibrium problems. 

The SIAM Hamiltonian is given by an impurity $H_\text{imp}$ that is coupled to a surrounding bath $H_\text{bath}$ by the 
hybridization $H_\text{hyb}$,
\begin{align}
	\label{eq:belMbath}
	H_\text{SIAM}&=H_\text{imp}+H_\text{bath}+H_\text{hyb},
	\\	
	H_\text{imp}&=-\mu \sum_\sigma n_{0\sigma} + U(t)\left(n_{0\uparrow}-\tfrac12\right)\left(n_{0\downarrow}-\tfrac12\right),\\
	H_\text{hyb}&=\sum_{p>0,\sigma}\left(V^\sigma_{0p}(t)\ads{0\sigma}\aas{p\sigma}+\text{h.c.}\right),
	\\
	H_\text{bath}&=\sum_{p>0,\sigma}(\epsilon_{p\sigma}(t)-\mu) \ads{p\sigma}\aas{p\sigma}.
\end{align}
Here the operator $\aas{p\sigma}$ ($\ads{p\sigma}$) annihilates (creates) an electron with spin $\sigma$ at bath site 
$p$ for $p>0$, and at the impurity for $p=0$. 

A mapping of the action (\ref{eq:dmftaction}) to the Hamiltonian (\ref{eq:belMbath}) requires that all impurity correlation 
functions are the same in the two models
\begin{multline}
\label{mapping1}
\frac{
\text{tr}_c\left(
\tcorder{\exP{S_\text{loc}}  \mathcal{O}(t_1) ...} 
\right)
}{
\text{tr}_c\left(
\tcorder{\exP{S_\text{loc}}}
\right)
}
\\
\stackrel{!}{=}
\frac{
\text{tr}_a\left(
\tcorder{\exP{-i\int\limits_C\dt H_\text{SIAM}(t)}  \mathcal{O}(t_1) ...} 
\right)
}{
\text{tr}_a\left(
\tcorder{\exP{-i\int\limits_C\dt H_\text{SIAM}(t)}}
\right)
}.
\end{multline}
Since the bath orbitals are noninteracting, the trace over the bath degrees of freedom can be performed analytically. The right hand side of~(\ref{mapping1}) then becomes
\begin{align}
\label{mapping2}
\frac{
\text{tr}_{a_0}\left(
\tcorder{\exP{S_\text{SIAM}}  \mathcal{O}(t_1) ...} 
\right)
}{
\text{tr}_{a_0}\left(
\tcorder{\exP{S_\text{SIAM}}}
\right)
},
\end{align}
where
\begin{multline}
\label{ssiamloc}
S_\text{SIAM}=-i\int\limits_C\dt H_\text{imp}(t)
\\
-i\int\limits_C\int\limits_C\dt_1\dt_2\sum_{\sigma}\Lambda^{\text{SIAM}}_{\sigma}(t_1,t_2)a_{0\sigma}^\dagger(t_1)a_{0\sigma}(t_2),
\end{multline}
with the discrete-bath hybridization function
\begin{align}
	\Lambda^\text{SIAM}_\sigma(t,t')&=\sum_{p}V^\sigma_{0p}(t)g_{p\sigma}(t,t')V^\sigma_{p0}(t'),
	\label{eq:actionreq}
\end{align}
and $g_{p}(t,t')$ is the Green function of an isolated bath site with energy $\epsilon_{p\sigma}(t)$,
\begin{align}
\label{gsignglebath}
g_{p\sigma}(t,t')
=
-i
\frac{
\text{tr}\left(
\tcorder{e^{-i\int_C ds (\epsilon_{p\sigma }(s)-\mu) a^\dagger a}  a(t)a^\dagger(t')} 
\right)
}{
\text{tr}\left(
\tcorder{e^{-i\int_C ds (\epsilon_{p\sigma }(s)-\mu)a^\dagger a}}
\right)
},
\end{align}
which evaluates to the 
result given by \eqeqref{eq:gsinglesite}.
The effective action (\ref{ssiamloc})  can be derived using coherent state path integrals, or equations of motions 
\cite{eck:09.09}. Because it is also a special case of the cavity expression (\ref{eq:cavity}), we shift the derivation 
to Sec.~\ref{sec:dmftaction}. 
In equilibrium, when all parameters of the impurity model are time-independent, 
\eqeqref{eq:actionreq}
reduces to the well-known expression~\cite{Georges96}
\begin{equation}
\label{mapping-eq}
-\frac{1}{\pi} \text{Im} \,
\Lambda^\text{SIAM}_\sigma (\omega+i0)
=
\sum_{p} 
|V^\sigma_{0p}|^2 
\delta(\omega+\mu-\epsilon_{p\sigma})
\end{equation}
for the spectral function of the bath.

We summarize this subsection by stating that the Hamiltonian (\ref{eq:belMbath}) is a valid representation of the 
DMFT action with hybridization function $\Lambda_\sigma(t,t')$ if parameters $V_{0p}^\sigma(t)$ and $\epsilon_{p\sigma}(t)$ can 
be chosen such that 
\begin{equation}\label{eq:gfrequirement}
\Lambda^\text{SIAM}_\sigma(t,t')=\Lambda_\sigma(t,t')
\end{equation}
on the entire contour $C$, i.e., for each of the five components given 
in Eq.~(\ref{gcomponent}), 
including the mixed 
components $\Lambda^ {\mixl}$ and $\Lambda^ {\mixr}$ which describe the correlations with the initial state. 
Due to the two time arguments, it is not immediately clear 
under which conditions a function $\Lambda_\sigma(t,t')$ can be represented in the form (\ref{eq:actionreq}) at all. 
We will thus first discuss the question of representability from a general 
perspective, before we attempt an explicit determination of the parameters $V$ and $\epsilon$ in Sec.~\ref{sec:bathfitting}.

In passing we note that we have restricted ourselves to a star-like layout of the impurity problem~(\ref{eq:belMbath}), 
i.e., there is no hopping between the bath orbitals. However, this is no limitation, since any bath with a more complicated 
geometry can always be mapped to a star geometry with the same effective action $S_\text{SIAM}$ by a suitable 
time-dependent unitary transformation (cf. Appendix~\ref{ap:role}).

\subsection{Representability}
\label{sec:mappingrepresentation}

Not every contour function with the symmetries (\ref{chermitian}) and the usual anti-periodic boundary conditions can be 
decomposed in the form~(\ref{eq:actionreq}). For equilibrium, the representation (\ref{eq:actionreq}) implies a positive 
spectral weight [cf.~\eqeqref{mapping-eq}], i.e., it relies on further analytic properties of the Green functions. The specification 
of analytical properties is less clear for  nonequilibrium Green functions. In the following we assert that representability in 
the form (\ref{eq:actionreq}) is nevertheless a general property of nonequilibrium Green functions: It follows for any 
(orbital-diagonal) Green function $G_{ii}(t,t')$ of a quantum system that evolves under the action of an arbitrary 
time-dependent Hamiltonian $H(t)$. 
To see this, we  expand $G(t,t')$ using eigenstates $|n\rangle$ and eigenenergies $E_n$ 
of the initial state Hamiltonian $H(0)$ (Lehmann representation). 
After some algebra, we get
\begin{multline}
\label{Lehmann1}
G_{ii}(t,t')
=
\sum_{nm}
\frac{e^{-\beta E_n}+e^{-\beta E_m}}{Z}
\\\times
W_{nm}(t)
W_{nm}^*(t')
g(E_{mn},t,t'),
\end{multline}
where $E_{mn}=E_{m}-E_{n}$, $g(E_{mn},t,t')$ is of the form~(\ref{g eps const}), and (operators with a hat are to be interpreted in the Heisenberg picture)
\begin{equation}
	\label{eq:wnm1}
	W_{nm}(t)=\left\{\begin{aligned}
		& e^{iE_{mn}t}\langle n | \hat c_i(t) | m \rangle  &&\text{ ~~~for real~}t\\
		&\langle n | \hat c_i(0) | m \rangle  &&\text{ ~~~for~}t=-i\tau
	\end{aligned}\right..
\end{equation}
This yields the representation~(\ref{eq:actionreq}) with one bath orbital for each 
of the pairs
$mn$ (the number of which is exponentially large in the system size), with time-independent orbital energy $\epsilon_{mn}=E_{mn}+\mu$ and hybridization 
$V_{0,mn}(t)=V_{mn,0}^*(t)=W_{nm}(t)\sqrt{(e^{-\beta E_n}+e^{-\beta E_m})/Z}$.

Similarly, it follows from the cavity expression (\ref{eq:cavity}) that the DMFT action 
is representable by a SIAM. 
To this end, we let $|n\rangle$ and $E_n$ denote eigenstates and eigenenergies of the cavity 
Hamiltonian $H^{(0)}(t)$ (i.e., the Hubbard Hamiltonian without lattice site $0$). Using a 
Lehmann representation of $G^{(0)}_{ij\sigma}(t,t')$, we again obtain a representation of 
the DMFT action by a SIAM with time-independent orbital energy $E_{mn}+\mu$, and hybridization 
matrix elements 
\begin{multline}
V_{0,mn}(t)=
\sqrt{\frac{e^{-\beta E_n}+e^{-\beta E_m}}{Z}}e^{iE_{mn}t}
\\\times
\sum_jt_{0j}(t)
\langle n | \hat c_j(t) | m \rangle.
\end{multline}

This concludes the proof that the DMFT action is representable by a SIAM. 
Of course, we have not shown that when DMFT is used as an approximation for finite-dimensional 
systems,  all solutions of the equations have a SIAM-representable DMFT bath.  
However, since representability relies on very general properties of physical 
Green functions as derived from the Lehmann representation, we will assume the existence of such a 
representation in the following
(i.e., we look only for DMFT solutions with this general property).

\section{The cavity method for nonequilibrium Green functions}
\label{sec:dmftaction}

Nonequilibrium DMFT equations and the effective impurity action can be derived using the cavity method,
by restating the arguments used for equilibrium DMFT \cite{Georges96}. In order to have a self-contained 
text, we now summarize the main steps, with technical details in Appendices~\ref{sec:genf} and~\ref{ap:selfcondarblatt}.
In equilibrium, the cavity method is typically formulated using Grassmann variables (e.g., Ref.~\onlinecite{vol:10}). 
However, in line with the notation used in the remainder of this paper, we keep using the equivalent language of 
contour-ordered expectation values.

We start from the grand-canonical partition function 
for a Hamiltonian $H(t)$, which can be either $H_\text{Hub}(t)$ or $H_\text{SIAM}(t)$,
\begin{equation}
	Z=\trace{\tcorder{\exP{-i\int\limits_CH(t)\dt }}}.
	\label{eq:part}
\end{equation}
The idea of the cavity method is to pick out one single site and trace out the remaining lattice
(for the SIAM, the isolated site will be the impurity).
We therefore separate the action into three parts,
\begin{align}
	\label{eq:actionS}
	S&=-i\int\limits_C\dt\,H(t)=S_0+\Delta S+S^{(0)},
\end{align}
with
\begin{align}
	\label{eq:dmftdefaction}
	S_0&=
      -i\int\limits_C\dt\biggl[U(t)(n_{0\uparrow}(t)-\tfrac12)(n_{0\downarrow}(t)-\tfrac12)
       \nonumber\\&\phantom{=-i\int\limits_C\dt\biggl[}\;\;
       -\mu\sum_\sigma n_{0\sigma}(t)\biggr],\\
	\Delta S&=-i\int\limits_C\dt\left[\sum_{i\neq 0,\sigma}t^\sigma_{i0}(t)\cds{i\sigma}(t)\ccs{0\sigma}(t)+\text{h.c.}\right],
        \nonumber\\
				\label{eq:hamexpansion}
	S^{(0)}&=-i\int\limits_C\dt\,H^{(0)}(t),
\end{align}
where $H^{(0)}$ is the Hamiltonian  $H(t)$ of the system ($H_\text{Hub}(t)$ or $H_\text{SIAM}(t)$) with site $0$ removed.
The Fock space of our system is given by the tensor product $\mathcal{F}=\mathcal{F}_{\text{rest}}\otimes\mathcal{F}_0$
with $\mathcal{F}_0$ being the Fock space of the isolated site and $\mathcal{F}_\text{rest}$ being the Fock space of 
the remaining sites. Corresponding to these subspaces 
the partial traces are
\begin{align}
	\traceb{O}&=\sum_{\{n_{i\sigma},i\neq0\}}\lwf{\{n_{i\sigma},i\neq0\}}O\rwf{\{n_{i\sigma},i\neq0\}},\nonumber\\
	\text{tr}_0({O})&=\sum_{\{n_{0\sigma}\}}\lwf{n_{0\sigma}}O\rwf{n_{0\sigma}},
\end{align}
where $\rwf{\{n_{i\sigma}\}}$ represents a state in the occupation
number basis.
We rewrite the partition function as
\begin{align}
	Z&=\text{tr}_0\left[\tcorder{\exP{S_0}\traceb{\exP{\Delta S+S^{(0)}}}}\right]
	\nonumber\\
        &=Z_{S^{(0)}}\text{tr}_0\left[\tcorder{\exP{S_0+\tilde{S}}}\right],
        \label{eq:part2}
\end{align}
where we defined
\begin{align}
	\label{eq:Stildedef}
	\exp(\tilde{S})&=\sum_{n=0}^\infty\frac{1}{n!}\est{(\Delta S)^n}_{S^{(0)}},
	\\
        \label{eq:S(0)def}
	\est{O(t)}_{S^{(0)}}&\equiv\frac{1}{Z_{S^{(0)}}}\traceb{\tcorder{\exP{S^{(0)}}O(t)}},\\
	Z_{S^{(0)}}&=\traceb{\tcorder{\exP{S^{(0)}}}}.\nonumber{}
\end{align}
Note that for $\exp(\tilde{S})$ there is still a contour ordering to be performed
[cf.~\eqeqref{eq:part2}], but one can 
perform the contour ordering on each
subspace independently
because the creation
(annihilation) operators that 
act on
$\mathcal{F}_\text{rest}$
anticommute with those acting on $\mathcal{F}_0$. It is thus possible
to calculate the trace over $\mathcal{F}_\text{rest}$ separately as
long as one keeps track of the correct sign.
This is done in Appendix~\ref{sec:genf}. The result is
\begin{multline}
	\tilde{S}=-i\sum_{n=1}^\infty\sum_{\sigma_1\dots\sigma_n'}\int\limits_C\dt_1\dots\int\limits_C\dt_n'\Lambda_{\sigma_1\dots\sigma_n'}(t_1,\dots,t_n')\\
        \times\cds{\sigma_1}(t_1)\dots\ccs{\sigma'_n}(t'_n),\label{eq:impurityaction}
\end{multline}
where we defined the $n$th-order hybridization functions
\begin{multline}
	\label{eq:deflam}
	\Lambda_{\sigma_1\dots\sigma_n'}(t_1,\dots,t_n')\equiv\frac{(-i)^{n-1}}{n!^2}\sum_{i_1,\dots,j_n}\!\!\!t_{0i_1}(t_1)\cdots t_{j_n0}(t_n')\\
	\times G^{(0),\text{c}}_{(i_1 \sigma_1),\dots,(j_n \sigma_n')}(t_1,\dots,t'_n),
\end{multline}
which involve connected cavity Green
functions $G^{(0),\text{c}}$ 
(obtained with $S^{(0)}$ only). 
The effective action is thus given by
\begin{align}
	S_\text{eff}&=S_0+\tilde{S},
        &
	Z_\text{eff}&=\frac{Z}{Z_{S^{(0)}}}=\text{tr}_0\left(\tcorder{\exp(S_\text{eff})}\right).
\end{align}
So far no approximation has been made, and the result is valid in general. However, it 
involves the higher-order hybridization functions~\eqref{eq:deflam}.
The latter are hard to compute in general, but they simplify
for the Hubbard model in $d$ $\to$ $\infty$ and for the
SIAM. In these cases the effective action is again of the
local form~\eqref{eq:dmftaction} with a noninteracting bath.

In the limit of infinite dimensions, with the quantum scaling~\cite{Metzner1989}
$t_{ij}\propto {d}^{-\mathcal{Z}_{ij}/2}$ 
($\mathcal{Z}_{ij}$  is the number of sites connected by hopping $t_{ij}$)  
one can one-to-one repeat the power counting arguments of the equilibrium 
formalism to show that the higher-order terms vanish, so that 
one is left with the quadratic bath part given by \eqeqref{eq:cavity} (cf.~Appendix~\ref{ap:selfcondarblatt}).

For a Bethe lattice with $\mathcal{Z}$ nearest neighbors, \eqeqref{eq:cavity} can be evaluated 
immediately. For neighboring sites $i,j$ of site $0$ we have $G^{(0)}_{ij}(t,t')\propto \delta_{ij}$ since there is no path that 
connects two distinct sites; such a path would involve site $0$, which has been removed. In the limit 
$\mathcal{Z}$ $\to$ $\infty$, we can further identify $G^{(0)}_{i\sigma}(t,t')=G_\sigma(t,t')$. The quantum 
scaling ensures that the summation over all nearest neighbors of $0$ stays finite.
This yields the action~\eqref{eq:dmftaction} with the hybridization~\eqref{bethe} 
(setting $t_{ij}(t)$ $=$ $v(t)/\sqrt{\mathcal{Z}}$ for nearest neighbors $i,j$), as previously derived in~Refs.~\onlinecite{eck:09.09,epjst2010a,Eckstein2010ramps}.

For a general lattice, one must formally express the cavity Green function in \eqeqref{eq:cavity}
in terms of the full lattice Green function in order to obtain an implicit relation of the form
$\Lambda=\Lambda[G]$, which can be achieved by utilizing the locality of the self-energy. It is
interesting to note that the DMFT equations also follow directly from the cavity method, if the
latter is applied to a generating functional of lattice Green functions. The locality of the
self-energy then need not be established by separate arguments, but it follows from the cavity
formalism itself. Because the precise form of the DMFT equations is not central to the mapping
problem, we present this self-contained derivation of the DMFT equations from the cavity method
in Appendix~\ref{ap:selfcondarblatt}.

One can also use the cavity argument to derive the effective action 
\eqref{ssiamloc} of a SIAM.
For a SIAM, the cavity Hamiltonian $H^{(0)}$ is noninteracting, so that all connected Green 
functions  $G^{(0),c}_{i_1\dots j_n}(t_1,\dots,t_n')$ vanish for $n\ge 2$. Thus only the first-order 
contribution remains in~\eqeqref{eq:deflam}. It takes the form~\eqref{eq:actionreq}, with 
 $G^{(0)}_{pp}(t,t')=g(\epsilon_p-\mu,t,t')$ [cf.~\eqeqref{g eps const}]. 
The Hubbard model on an infinite-dimensional lattice and the SIAM thus have the same
local action, provided their hybridization functions are the same [cf.~\eqref{eq:gfrequirement}]. 
The bath parameters of the SIAM for a given hybridization function are determined in the next section.

\section{Construction of the SIAM representation for nonequilibrium Green functions}
\label{sec:bathfitting}

According to Sec.~\ref{sec:mappingrepresentation}, we may assume that the hybridization function $\Lambda_\sigma(t,t')$
has a valid representation in the form~(\ref{eq:actionreq}). In this section, we explicitly construct the parameters 
$V^\sigma_{0p}(t)$ and $\epsilon_{p\sigma}$ for a given $\Lambda_\sigma(t,t')$. In the course of this, we will introduce two distinct baths, denoted \emph{first} ($-$) and \emph{second} ($+$),
which take care of the fact that in nonequilibrium initial correlations and the dynamic built-up of correlations must be represented differently.
While the first bath involves sites that are coupled to the system already at $t=0$ and usually vanishes as time proceeds, the second one couples additional sites for $t>0$ and typically exists even in the final (steady) state for $t$ $\to$ $\infty$.

\subsection{Memory of the initial state (the first bath)}
\label{sec:memofinit}

To determine the free parameters of the SIAM explicitly
we follow Ref.~\onlinecite{eck:09.09} and 
assume that the bath states can be characterized by a continuous energy band with density of states 
$\rho_{-}(\epsilon) = \sum_p \delta(\epsilon-\epsilon_p(0))$  and hybridization $V_{-}(\epsilon,t)$
(for simplicity we drop the spin index; the subscript ``$-$'' adverts to the ``first bath''). 
We also assume that the site energies $\epsilon_{p}$ are time-independent, buth this is no restriction because 
a SIAM with time-dependent $\epsilon_{p}(t)$ and $V_{0p}(t)$ has the same effective action (\ref{ssiamloc}) like one with
time-independent energies $\epsilon_{p}'=\epsilon_p(0)$ and modified hoppings $V_{0p}'(t) = V_{0p}(t)\exP{-i\int_0^t ds 
[\epsilon_p(s)-\epsilon_p(0)]}$ [cf.~\eqref{eq:gsinglesite} and~\eqref{eq:actionreq}]. We thus require a 
representation of $\Lambda(t,t')$ 
in the form 
\begin{align}
	\label{eq:ansatzlambda}
	\Lambda_{-}(t,t')=\int\limits_{-\infty}^\infty\text{d}\epsilon\,\rho_{-}(\epsilon) V_{-}(\epsilon,t)g(\epsilon-\mu,t,t')V_{-}(\epsilon,t')^*.
\end{align}
In the following, also $\rho_{-}(\epsilon)$ is absorbed in the hybridization matrix elements, i.e., the bath density of states 
is chosen to be constant. 

It is important to recall that $V_{-}(\epsilon,t)$ 
must be
constant on the imaginary time 
branch, i.e., $V_{-}(\epsilon,-i\tau)=V_{-}(\epsilon,0)$, since it will be used as hopping parameter in the effective SIAM.
To determine its time dependence we consider the mixed components
[cf.~\eqeqref{g eps const}],
\begin{align}
	\label{eq:followme1}
	\Lambda_-^\mixr(t,\tau)
	&=i\int\limits_{-\infty}^\infty \text{d}\epsilon V_{-}(\epsilon,t)f(\epsilon-\mu)\text{e}^{-i(\epsilon-\mu)(t-i\tau)}V_{-}(\epsilon,0)^*
	\\
	&\stackrel{!}{=} \Lambda^ {\mixr}(t,\tau).
	\label{fix lminus}
\end{align}
To solve~\eqref{fix lminus} for the parameters $V_-(\epsilon,t)$, we make use of the analytical properties 
of the mixed Green functions (see Appendix~\ref{sec:anaprop}): In analogy to the Matsubara Green function, one can 
introduce a Fourier series with respect to Matsubara frequencies, and continue the result to a function 
$\Lambda^ {\mixr}(t,z)$ which is analytic on the upper and lower complex frequency plane; $\Lambda^ {\mixr}(t,\tau)$ 
is then uniquely determined by the generalized spectral function [cf.~\eqref{eq:tspectral}]
\begin{equation}
C^\mixr(t,\epsilon)=\frac{1}{2\pi} \left(\Lambda^\mixr(t,\epsilon+i0)-\Lambda^\mixr(t,\epsilon-i0) \right).
\end{equation}
For time $t=0$, the spectral function coincides with the real and positive equilibrium spectral function 
\begin{equation}
C(\epsilon)=\frac{i}{2\pi} \left(\Lambda^M(\epsilon+i0)-\Lambda^M(\epsilon-i0) \right),
\end{equation}
see Appendix~\ref{sec:anaprop} for details.

When we Fourier transform both sides of~\eqref{eq:followme1}  to Matsubara frequencies and analytically 
continue the result to real frequencies, 
we obtain
\begin{align}
	C^\mixr(t,\epsilon)&=\exP{-i\epsilon t}V_{-}(\epsilon+\mu,0)^* V_{-}(\epsilon+\mu,t).
\end{align}
This allows us to obtain an explicit expression for the hopping matrix elements,
\begin{align}
\label{eq:mixed 0 }
V_{-}(\epsilon+\mu,0)&=\sqrt{C(\epsilon)},
\\
V_{-}(\epsilon+\mu,t)&=\frac{\exP{i\epsilon t}C^\mixr(t,\epsilon)}{V_{-}(\epsilon+\mu,0)^*}.
\label{eq:mixed}
\end{align}
Note that a phase factor of $V_{-}(\epsilon+\mu,0)$ can be chosen freely, without changing the resulting action.

The mixed component therefore already fixes all free parameters $V_{-}(\epsilon,t)$, 
and thus also $\Lambda_{-}(t,t')$ {\em on the entire contour C}.
If $t$ or $t'$ is an imaginary time, $\Lambda_{-}(t,t')$ coincides with $\Lambda(t,t')$
by construction.
However, 
for real times $t,t'$ the difference
\begin{equation}
	\label{eq:bathsplit}
	\Lambda_{+}(t,t')\equiv\Lambda(t,t')-\Lambda_{-}(t,t')
\end{equation}
is in general not equal to zero.
For a system out of equilibrium the correlations between the initial state and states at $t=\infty$ 
usually vanish, i.e., \makebox{$\Lambda^\mixr(t,\tau)\stackrel{t\to\infty}{\rightarrow}0$}.
Hence one would expect~\eqref{eq:mixed} to vanish (which is also found numerically).
For the final state to be nontrivial, $\Lambda_{+}(t,t')$ therefore has to be nonzero. The 
bath
$V_{-}(t)$ is thus not sufficient to
represent $\Lambda$.
For this reason, we introduce a ``second bath'' with hoppings 
$V_{+}(t)$ 
which represent the contribution $\Lambda_{+}(t,t')$.
The Weiss field $\Lambda_{-}(t,t')$ can then be understood to describe the fading memory of the 
initial state, while 
$\Lambda_{+}(t,t')$ is building up to describe the steady state for $t=\infty$. 
In the following we refer to $\Lambda_{-}(t,t')$ as the \emph{first} and to $\Lambda_{+}(t,t')$ as 
the \emph{second} Weiss field, respectively.

In passing we note that it is usually an ill-conditioned problem to determine the generalized
spectral function $C^\mixr(t,\epsilon)$ for a given $\Lambda(t,t')$. However, when one solves
the DMFT equations using the exact time propagation of the SIAM Hamiltonian, one has 
direct access to the real-frequency representation of all impurity Green functions via the 
Lehmann expressions given in Appendix~\ref{sec:anaprop}. 
Because also the
Keldysh-Kadanoff-Baym equations for a general DMFT self-consistency can be formulated in terms of the time-dependent 
spectral functions instead of imaginary-time quantities, as described in detail in Ref.~\onlinecite{Eckstein2010},
analytical continuation is no problem.

\subsection{The equilibrium case}
It is assuring to verify that for a system in equilibrium the second bath contribution indeed vanishes, such that 
the system is completely described by the first Weiss field $\Lambda_{-}(t,t')$ or the first bath $V_{-}(\epsilon,t)$ respectively.
In equilibrium, $\Lambda$ is entirely determined by its spectral function
\begin{align}
  \label{lam spec eq} 
	\Lambda(t,t')&=i\int\limits_{-\infty}^\infty\text{d}\epsilon\, C(\epsilon)\left(f(\epsilon)-\Theta_C(t,t')\right)\text{e}^{-i\epsilon(t-t')}.
\end{align}
On the other hand, we have $C^\mixr(t,\epsilon)=\exP{-i\epsilon t}C(\epsilon)$ [cf.~\eqref{eq:trivtime}] and 
thus find a time-independent coupling
$V_{-}(\epsilon+\mu,t)=\sqrt{C(\epsilon)}$
from \eqeqref{eq:mixed}.
When this is reinserted in~\eqref{eq:ansatzlambda}, we recover~\eqref{lam spec eq}.

\subsection{The second bath}
\label{sec:zerotempbath}
The construction of a hybridization for the second Weiss field,
\begin{equation}
\label{lamda plus 1}
\Lambda_+(t,t')
\stackrel{!}{=}
\sum_{p} V^+_{0p}(t)g_{p\sigma}(t,t')V^+_{p0}(t'),
\end{equation}
is quite different to the previous discussion for $\Lambda_-(t,t')$. 
From its definition~\eqref{eq:bathsplit} it follows that the imaginary-time components  $\Lambda_+^M$, $\Lambda_+^ {\mixr}$,
and $\Lambda_+^ {\mixl}$ vanish, and thus we set $V^+_{0p}(t=0)=0$ for all bath sites $p$ representing $\Lambda_+$. 
Furthermore, it is convenient to choose a simple time dependence for the bath energies $\epsilon_p(t)$, which will 
take a value $\epsilon_p(0)$ in the initial state, and a different but time-independent value $\epsilon_p(\infty)$ for times $t>0$. 
As discussed above~\eqref{eq:ansatzlambda}, the value of $\epsilon_p(t)$ can be chosen freely for $t>0$, because any 
time dependence can be absorbed in the time dependence of the hoppings 
$V^+_{0p}$. 
The  initial-state value $\epsilon_p(0)$, on 
the other hand, enters~\eqref{lamda plus 1} only via the occupation functions $f(\epsilon_{p}-\mu)$ 
(in $\Lambda_+^<$) and $f(-\epsilon_{p}+\mu)$ (in $\Lambda_+^>$). Hence we make a simple choice and take $f(\epsilon_{p}-\mu)$ 
to be either $0$ or $1$. In summary, we attempt to  represent $\Lambda_+$ by a bath with two sets of orbitals 
$B_\text{occ}$ and $B_\text{empty}$, such that 
\begin{subequations}
\label{eq:decomp2}
\begin{align}
	-i\Lambda^{<}_+(t,t')&=\sum_{p \in B_\text{occ}}  V_{0p}^+(t) V_{0p}^+(t')^*,
	\\
	i\Lambda^{>}_+(t,t')&=\sum_{p \in B_\text{empty}}  V_{0p}^+(t) V_{0p}^+(t')^*.
\end{align}
\end{subequations}
In contrast 
to \eqeqref{fix lminus} 
for the first bath, these equations have the form of a standard matrix decomposition. When the system is particle-hole symmetric, i.e., $\mu$ $=$ $0$ for the Hubbard model~\eqref{eq:hubbard}, we have 
$\Lambda^<_+(t,t')=\Lambda^>_+(t,t')^*$, which is satisfied when occupied and unoccupied bath orbitals come in pairs 
with complex conjugate hoppings. It is then sufficient to solve one of the two equations.
Note that the form of the decomposition~\eqref{eq:decomp2} requires $-i\Lambda^<_+$ and $i\Lambda^>_+$ to 
be positive definite matrices. In Appendix~\ref{ap:pdlambdap} we establish this property 
under the general assumption that the original $\Lambda$ is representable by a SIAM.

For a numerical implementation 
of \eqeqref{eq:decomp2} 
we discretize the time $t$. 
With $t_n\equiv n\times\Delta t\in[0,N\times \Delta t=t_\text{max}]$ we have
	\begin{equation}
		\label{eq:secondWeissDis}
		(-i\Lambda^<_+)_{nn'}\equiv -i\Lambda^<_+(t_n, t_{n'})=
		\sum_{p}  
		V^+_{0p}(t_n) V^+_{0p}(t_{n'})^*,
	\end{equation}
	where $p\geq1$ runs over the initially occupied bath sites in $B_\text{occ}$
(the equation for $\Lambda_+^>$ is treated analogously).
To solve the equation, one may use an eigenvector decomposition of the matrix $(-i\Lambda^<_+)$,
\begin{align}
&(-i\Lambda^<_+)_{nn'}=\sum_{p=1}^{N} U_{np} \,a_p \,U^*_{n'p}
\end{align}
with a unitary matrix $U_{np}$ and, since $-i\Lambda^<_+$ is positive definite, only positive eigenvalues $a_p$.
We can thus identify time-dependent hopping matrix elements
\begin{align}
V^+_{0p}(t_n) \equiv U_{np}\sqrt{a}_{p}.
\end{align}
We emphasize that, in general, the number of bath sites needed for the representation equals the number of timesteps in the discretization.

\begin{figure}[!t]
\includegraphics[width=0.80\columnwidth]{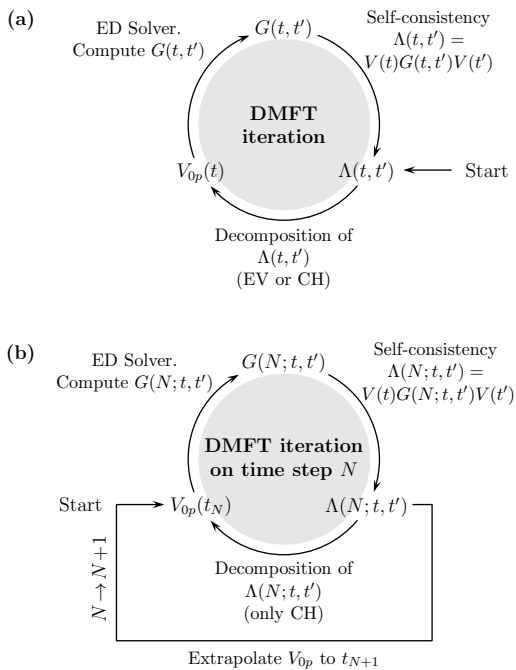}
\caption{Possible time-evolution schemes. (a) DMFT iteration involving all times $t,t'\leq t_\text{max}$ simultaneously, (b) progressive time-propagation scheme for which the DMFT self-consistency needs to be established only on the current timestep $N$ at a time. In panel (b), the notation $G(N;t,t')$ indicates that $t$ and $t'$ belong to the time slice $N$, i.e., one time (either $t$ or $t'$) is identical to $N\times\Delta t$ and the other is smaller or equal to $N\times\Delta t$.}
\label{fig:dmft_schemes}
\end{figure}

Although this approach is rather straightforward, it has a slight conceptual disadvantage: 
When the hoppings $V_{p}(t_n)$ are recomputed for a larger maximum time $t_\text{max}$, 
their value will be modified on {\em all} previous times. Thus the eigenvector decomposition 
cannot be used within a time-propagation scheme in which the DMFT solution is computed 
by successively extending $t_\text{max}$ timestep by timestep. Instead, one would have 
to perform a DMFT iteration as indicated in Fig.~\ref{fig:dmft_schemes}a, i.e., starting from a suitable guess for $G$ and in turn for $\Lambda$,
\begin{enumerate}
 \item [(i)] one computes the parameters $V_{0p}(t_n)$ {\em for all times} 
$t \le t_\text{max}$, 
 \item [(ii)] one computes the Green function~\eqref{dmft g} {\em for all times} 
$t,t' \le t_\text{max}$,
 \item [(iii)] one solves the DMFT self-consistency to get $\Lambda$ {\em for 
all times},
\end{enumerate}
and 
one  
iterates steps (i)-(iii) until convergence. Compared to a time-propagation 
scheme which updates the functions only on one time slice~\cite{Eckstein2010}
when going from $t_\text{max}$ to $t_\text{max}+\Delta t$, an iterative scheme thus requires considerably more computing resources.

To resolve this problem we employ a Cholesky decomposition, which has the property that 
parameters $V(t_n)$ can be determined independently of the parameters $V(t)$ for $t>t_n$.
Explicitly, the Cholesky decomposition 
of the $N\times N$ positive definite hermitian matrix
\begin{align}
\label{eq:choltriang}
\begin{pmatrix}
-i\Lambda^<
\end{pmatrix}
\!=\!\begin{pmatrix}
V_{00} &  &   & 
\\
V_{10} & V_{11} &   & 
\\
\vdots&\vdots& \ddots &
\\
V_{N0} & V_{N1} & \cdots & V_{NN} 
\end{pmatrix}\!\!
\begin{pmatrix}
V_{00}^* & V_{10}^*   & \cdots & V_{N0}^*
\\
&  V_{11}^*  &  \cdots & V_{N1}^*
\\
&& \ddots &\vdots
\\
 &  &  & V_{NN}^* 
\end{pmatrix}
\end{align}
in terms of triangular matrices $V$ and $V^\dagger$ is given by
\begin{align}
	V_{nn'} &= \frac{1}{V_{n'n'}} \left[ (-i\Lambda_+^<)_{nn'} - \sum_{m=1}^{n'-1} V_{nm}V_{n'm}^* \right],~(n'<n),\nonumber\\ 
	V_{nn} &= \sqrt{ (-i\Lambda_+^<)_{nn} - \sum_{m=1}^{n} |V_{nm}|^2 },
\label{eq:Vdisc}
\end{align}
and $V_{nn'}$ $=$ $0$ for $n'$ $>$ $n$.
We then choose
\begin{equation}
V_{np}=V_{0p}^+(t_n),
\end{equation}
i.e., the $p$-th column of $V$ yields the time-dependent hybridization to bath orbital $p$.
The triangular structure of $V$  in~Eq.~(\ref{eq:choltriang}) implies that a new bath orbital is coupled to the system at 
each timestep, and it allows for the recursive determination of $V$ in~\eqref{eq:Vdisc},
which works line-by-line, i.e., by timestep by timestep. The associated time-propagation scheme is sketched in Fig.~\ref{fig:dmft_schemes}b. 
Here, the two-time quantities $G$ and $\Lambda$ are updated not as a whole but on the time slice $N$ only (in the figure the notation $G(N;t,t')$ indicates 
that either $t$ or $t'$ is equal to the current maximum time $t_\text{max}=N\times\Delta t$). Furthermore, the extrapolation of the hopping matrix 
elements $V_{0p}(t)$ with $t\leq N\times\Delta t$ to the next timestep $N+1$ ensures, together with a small $\Delta t$, that the DMFT self-consistency 
is reached within a few (typically one to three) cycles. It is this fact which makes the time-propagation scheme more efficient than the DMFT 
iteration scheme [cf.~Fig.~\ref{fig:dmft_schemes}a].

\section{Approximate representations}
\label{sec:approx_repres}
\label{sec:approxrepr}
It is clear that a meaningful approximate solution 
to~\eqeqref{eq:secondWeissDis} 
is required in practice since a numerical treatment of $H_\text{SIAM}(t)$ is limited to a small 
number of sites. 
In this section we will develop approximation schemes to the Cholesky and the eigenvector decomposition that yield a decomposition of the form
\begin{equation}
	(-i\Lambda^<_+)_{nn'}\approx\sum_{p=1}^L V_{np}V^*_{n'p},
\end{equation}
(and similar for $i\Lambda^>_+$),
where $L\ll N$ is a fixed finite number which is equal to the rank of $V$ and thus to 
the rank of the approximate $-i\Lambda^<_{+,\text{approx}}=V V^\dagger$. 
To represent $\Lambda_+$ by a finite number of sites $L_\text{bath}$, we need to find low-rank approximations 
for $-i\Lambda^<_+$ and $i\Lambda^>_+$.
We discuss this in detail only for  $-i\Lambda^<_+$.
\subsection{Low-rank Cholesky approximation}
\begin{figure*}[!t]
{\includegraphics[width=\linewidth]{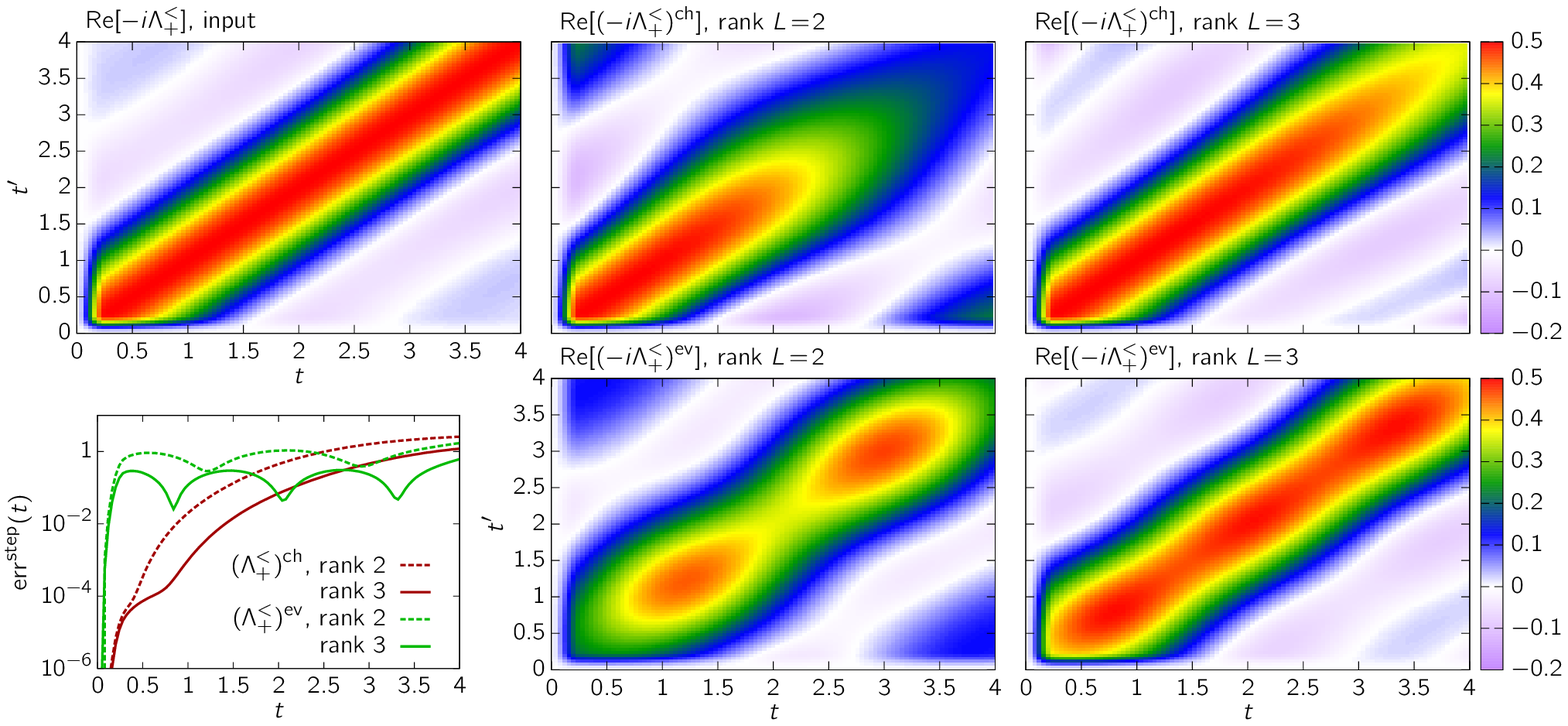}}\centering
\caption{Comparison of low-rank Cholesky ($(-i\Lambda_+^<)^\text{ch}$) and eigenvector ($(-i\Lambda_+^<)^\text{ev}$) approximation. The top panel on the left shows the real part of an typical 
input Weiss field $-i\Lambda_+^<$ (obtained from a calculation with six bath sites, i.e., $L_\text{bath}=2L=6$ according to the setup described in Sec.~\ref{sec:bethequench} with $U=2$). 
The Weiss field in the top panel of the second (third) column displays the an approximate hybridization $V^\text{ch}_{np}$ with rank $L=2$ ($L=3$) that was calculated with the low-rank Cholesky approach. In the lower panel of the second (third) column an approximate Weiss field obtained from a rank $L=2$ ($L=3$) eigenvector decomposition is shown. The bottom left panel compares the stepwise error of both approximations as defined in \protect\eqref{eq:stepwiseerror}.}
\label{fig:lambdaappr}%
\end{figure*}%
\label{sec:lowrank}
A finite rank $L$ for the hybridization matrix $V_{np}=V^+_{0p}(t_n)$ can be realized with the ansatz
\begin{align}
V^\text{ch}\!=\!\left(\begin{matrix} V^\text{ch}_{11} & 0 & \hdotsfor{3} \\
	\vdots & \ddots & 0 & \hdotsfor{2}\\
	V^\text{ch}_{L1} & \dots & V^\text{ch}_{LL} & 0 & \hdots\\
	V^\text{ch}_{L+1,1} & \dots & V^\text{ch}_{L+1,L} & 0 & \hdots\\
	\vdots & \vdots & \vdots & \vdots & \vdots
\end{matrix}\right)
\!\equiv\!
\left(\begin{matrix} (Q^s)^\dagger & 0 & \dots \\
	(\vec{q}_{s+1})^\dagger & 0 & \dots \\
	(\vec{q}_{s+2})^\dagger & 0 & \dots \\
	\vdots & \vdots & \vdots
\end{matrix}\right)\!,
\end{align}
for $s$ $\ge$ $L$,
where we introduce abbreviations
$Q^{s}_{ij}$ $=$ ${V_{ji}^\text{ch}}^*$ for $j$ $\le$ $L$, $i$ $\le$ $s$ and $\vec{q}_{s+n}$ $=$ $(V^\text{ch}_{s+n,1},\dots,V^\text{ch}_{s+n,L})^\dagger$.
We further define the corresponding approximate Weiss field as
\begin{equation}
	(-i\Lambda_+^<)^\text{ch}_{nn'}\equiv \sum_{p=1}^L V_{np}^\text{ch}(V^\text{ch}_{n'p})^*\stackrel{!}{\approx} (-i\Lambda_+^<)_{nn'}.
\end{equation}
During the first $L$ timesteps we have enough free parameters $V^\text{ch}_{np}$ for an exact matrix decomposition of $-i\Lambda_+^<$ and can thus rely on the Cholesky decomposition to fill up the matrix $Q^L$. After that we have to perform approximate updates of the hybridization. Let us assume that we already performed $s\ge L$ steps and found the approximation $Q^s$. We denote the corresponding exact second Weiss field with $(-i\Lambda_+^<)_s$. In the next timestep, the matrix $(-i\Lambda_+^<)_s$ gets updated by one line and column $\vec{a}_{s+1}\equiv((-i\Lambda_+^<)_{s+1,1},\dots,(-i\Lambda_+^<)_{s+1,s})^\dagger$. To find an approximation for the second bath, we have to minimize the error of 
\begin{align}
\left(\begin{matrix} (Q^s)^\dagger & 0 \\
	(\vec{q}_{s+1})^\dagger & 0 
\end{matrix}\right)
\left(\begin{matrix} Q^s & \vec{q}_{s+1} \\
	0 & 0 
\end{matrix}\right)\!&=\!
\left(\begin{matrix} (Q^{s})^\dagger Q^s & (Q^s)^\dagger \vec{q}_{s+1} \\
	(\vec{q}_{s+1})^\dagger Q^s & (\vec{q}_{s+1})^\dagger\vec{q}_{s+1}
\end{matrix}\right)\nonumber\\
\!&\approx\!
\left(\begin{matrix} (-i\Lambda_+^<)_s & \vec{a}_{s+1} \\
	\vec{a}^\dagger_{s+1} & (-i\Lambda_+^<)_{s+1,s+1} 
\end{matrix}\right)\!.
\end{align}
Since we know that $(Q^s)^\dagger Q^s\approx (-i\Lambda_+^<)_s$ is a good approximation, we only update the new components by minimizing
\begin{align}
	\label{eq:minF}
	\text{min } F(\vec{q}_{s+1})&=2|| Q \vec{q}_{s+1}-\vec{a}_{s+1} ||^2\\
	&\phantom{=}\,\,+ \abs{(\vec{q}_{s+1})^\dagger\vec{q}_{s+1}-(-i\Lambda_+^<)_{s+1,s+1}}^2,\nonumber
\end{align}
with respect to $\vec{q}_{s+1}$. ``$||\cdot||$'' is the usual Euclidean norm. The optimal $\vec{q}_{s+1}$ can then be used as approximate update for the hybridization matrix and the same procedure can be repeated. We emphasize that an update of the Weiss field $t_\text{max}\rightarrow t_\text{max}+\Delta t$, i.e., $s\rightarrow s+1$, leaves all previously calculated entries of $V^\text{ch}$ unaltered. Furthermore, for $n,p$ $\le$ $L$, the matrix $V^\text{ch}_{np}$ is equal to the exact Cholesky decomposition.

In Fig.~\ref{fig:lambdaappr} we show typical 
results for the low-rank Cholesky approximation. The corresponding input Weiss field is given for $100$ timesteps with $\Delta t=0.04$ 
and is shown in the top left panel. While an exact Cholesky decomposition would thus require a rank $L=100$ hybridization, the low-rank approach gives a reasonable approximation with a much smaller rank. For $L=2$ the approximate Weiss field $(-i\Lambda_+^<)^\text{ch}$ shows a good agreement with the input data up to times $t=1.5$ while the $L=3$ approximation allows to describe times up to $t=2.5$. Note however, that the quality of the approximation also depends on the exact form of the input Weiss field.

For larger times we find that the approximate Weiss field becomes less and less accurate. This is readily understood by viewing
\begin{equation}
	(-i\Lambda_+^<)^\text{ch}(t,t')= \sum_p V^\text{ch}_{0p}(t)V^\text{ch}_{0p}(t') ^*\equiv \vec{v}(t') ^\dagger\vec{v}(t)
\end{equation}
as a scalar product with $(\vec{v}(t))_p=V_{0p}^\text{ch}(t)$
in an $L$-dimensional vector space. 
Since the off-diagonal elements of $-i\Lambda_+^<$ are small, we need $\vec{v}(t') ^\dagger\vec{v}(t)\approx 0$ for $t\gg t'$. However, 
our approximation relies on a fixed dimension $L$, 
and it thus runs out of orthogonal 
vectors after some time. The only remaining way to approximate small off-diagonal values of $-i\Lambda_+^<$ is 
to reduce $||\vec{v}(t)||^2 = -i\Lambda_+^<(t,t)$.
However, this automatically leads to a decay of the  diagonal elements, as observed in Fig.~\ref{fig:lambdaappr}.

For completeness we also show results for the hybridization $V_{0p}^\text{ch}(t)$ in the top panel of Fig.~\ref{fig:hybrid}. 
As an important property we note that the Cholesky approach yields a hybridization that 
is a continuous function of time,
which is an important requirement for it to be used as hopping parameter in the SIAM Hamiltonian. 
Because of $\Lambda_+(0,t)=\Lambda_+(t,0)=0$ we obtain $V_{0p}^\text{ch}(0)=0$.

\subsection{Low-rank eigenvector decomposition}
\label{sec:eigred}
A different approach to find an optimal low-rank approximation for a given rank $L$ uses an eigenvector decomposition. We assume that the eigenvalues $a_p$ in the following decomposition are ordered in magnitude, so that the largest eigenvalue is given by $a_1$ and the smallest one by $a_{N}$. Then
\begin{align}
(-i\Lambda_+^<)_{nn'}&=\sum_{p=1}^{N} U_{np}a_p U^*_{n'p}\approx\sum_{p=1}^{L} (U_{np}\sqrt{a_p})(\sqrt{a_p} U^*_{n'p})\nonumber\\
&=\sum_{p=1}^L V^\text{ev}_{np}(V^\text{ev}_{n'p})^*\equiv (-i\Lambda_+^<)^\text{ev}_{nn'},
\end{align}
where $V^\text{ev}_{np}\equiv U_{np}\sqrt{a}_{p}$. This approximation minimizes the error with respect to the spectral norm, i.e.,
\begin{figure}[!t]
	{\includegraphics[width=\linewidth]{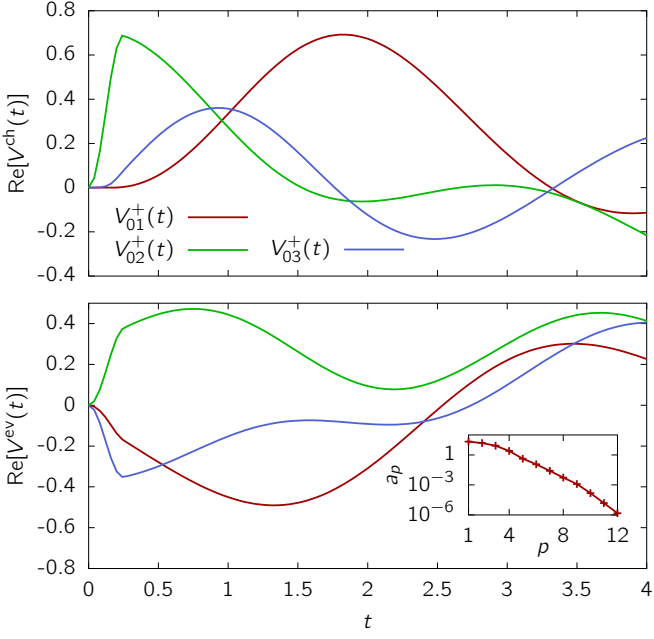}}\centering
	\caption{Time evolution of the hybridization $V^+_{0p}(t)$. Both panels show a rank $L=3$ approximation that corresponds to the input Weiss field $-i\Lambda^<_+$ displayed in Fig. \ref{fig:lambdaappr}. To obtain $V_{0p}^\text{ch}(t)$ in the top panel we used the low-rank Cholesky approach. In the lower panel we plot the hybridization $V_{0p}^\text{ev}(t)$, which was calculated using the low-rank eigenvector approximation. The inset shows the decay of the eigenvalues $a_p$ of $-i\Lambda_+^<$. }
\label{fig:hybrid}%
\end{figure}%
\begin{equation}
	||\Lambda_+^<-(\Lambda_+^<)^\text{ev}||_\text{mat}\equiv\mathop{\text{max}}_{||\vec{x}||=1} ||\,[\Lambda_+^<- (\Lambda_+^<)^\text{ev}]\,\,\vec{x}\,\,||,
\end{equation}
where $\vec{x}$ is a vector in the corresponding vector space and $||\cdot||$ the Euclidean norm. For our approximation the norm yields $a_{L+1}$. 

The approximation is best suited for matrices with an eigenvalue spectrum that drops off rapidly. As an example we consider again the 
input data shown in Fig.~\ref{fig:lambdaappr}. Indeed we find that the eigenvalues $a_p$ of $-i\Lambda_+^<$ decrease fast, cf. the inset in the lower panel of Fig.~\ref{fig:hybrid}. In the lower panel of the second (third) column of Fig.~\ref{fig:lambdaappr} the corresponding rank $L=2$ (rank $L=3$) approximation of $-i\Lambda_+^<$ is shown. In contrast to the Cholesky approach we find that no special attention is paid to small times. This is due to the fact that many eigenvectors are discarded which affects the whole matrix. As an important consequence, the error is spread over the resulting approximate matrix, as discussed in the next subsection.

Similarly to the Cholesky approach, the eigenvector approximation yields a continuous hybridization, cf. the lower panel in Fig.~\ref{fig:hybrid}.
We again find $V^\text{ev}_{0p}(0)=0$ as required by $\Lambda_+(0,t)=\Lambda_+(t,0)=0$.

\subsection{Comparison of Cholesky and eigenvector approximation}
\label{sec:cmpcholev1}
We investigate how the error of both approximations is spread over the resulting Weiss field. This is most easily understood by looking at the stepwise error at time $\tau$ which we define as
\begin{equation}
	\label{eq:stepwiseerror}
\text{err}^\text{step}(A,\tau)=\sqrt{\sum_{n=1}^{N}(2-\delta_{nN})\abs{(\Lambda_+^<)_{nN} - A_{nN}}^2},
\end{equation}
with $N=\tau/\Delta t$. Numerical results for the 
input Weiss field $-i\Lambda_+^<$ shown in Fig.~\ref{fig:lambdaappr} are plotted in the bottom left panel of the same figure. As expected we find a monotonic increase of the error, which is very small for short times, for the Cholesky approach. For the eigenvector approximation, on the other hand, the error is spread almost equally over the whole matrix. This allows for a better approximation of $-i\Lambda_+^<$ as a whole, e.g., $\text{err}[(\Lambda_+^<)^\text{ev}]=0.09$ compared to $\text{err}[(\Lambda_+^<)^\text{ch}]=0.17$ for the rank $L=3$ approximation of the 
input data in Fig.~\ref{fig:lambdaappr}, with
\begin{align}
	\text{err}[A]\equiv \frac{||\Lambda_+^<-A||}{||\Lambda_+^<||},\quad ||A||\equiv \sqrt{\sum_{nn'} |A_{nn'}|^2}.
\end{align}
However, the maximum time that can be represented accurately using the eigenvector decomposition
is not known beforehand, which can pose a problem if we use it within the DMFT self-consistency cycle
(cf.~following Section)

\section{Numerical results}
\label{sec:bethequench}

\subsection{Setup}

In this part, we apply the method developed in Sec.~\ref{sec:dmftaction}-\ref{sec:approx_repres} to a
simple test system. We study the time evolution for a 
Hubbard model on the Bethe lattice 
in the limit of infinite coordination number $\mathcal{Z}$ with 
time dependent nearest-neighbor $t_{ij}$ $=$ $v(t)/\sqrt{\mathcal{Z}}$  and constant on-site interaction $U$ [cf.~\eqeqref{eq:hubbard}]. 
We start from the atomic limit ($v$ $=$ $0$)
and smoothly but rapidly turn on the hopping up to a final value of $v$ $=$ $v_0$ $\equiv$ $1$ at time $t_{1}>0$. 
All quantities below are thus measured in units of $v_0$.
For the ramp of the hopping we choose a cosine-shaped time dependence,
\begin{align}
\label{eq:ramp}
v(t)=\begin{cases}
		\tfrac{1}{2}(1-\cos(\omega_0 t))&\text{for~}t<t_{1},~\omega_0=\tfrac{\pi}{t_{1}}\\
		1&\text{for~}t\ge t_{1},
\end{cases}
\end{align}

The initial state is assumed to be at zero temperature in the paramagnetic phase at half filling.
The paramagnetic DMFT solution for  $U>0$, $v=0$, and $T=0$ corresponds to a spin-disordered 
state with entropy $\ln(2)$ per lattice site
($\ket{\uparrow}_i$ and $\ket{\downarrow}_i$ are degenerate on each lattice site $i$), which 
is equivalent of taking the limit $v\to0$ and $T\to0$ such that the 
temperature is always larger than the Neel temperature $T_\text{Neel} \propto v^2/U$. 
The particle density is $\langle n\rangle=1$ since we have $\langle n_{i\sigma}\rangle=\tfrac{1}{2}$ 
for all lattice sites. We further have zero double occupation independent of the (positive) value of $U$.

\begin{figure}[!t]
\includegraphics[width=0.70\columnwidth]{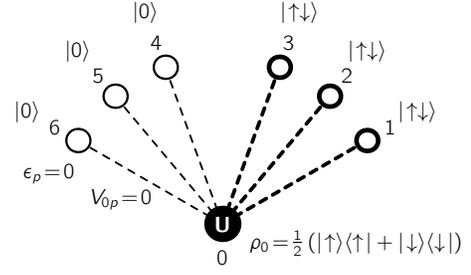}
\caption{Atomic limit: SIAM representation of the DMFT initial state for the paramagnetic phase 
of the Bethe lattice and $L_\text{bath}=6$. While the full dot denotes the impurity site with on-site 
Coulomb repulsion $U$, the open dots mark the bath with $V_{0p}(0)=0$ in the initial state ($1\leq p\leq L_\text{bath}$).}
\label{fig:initialstate}
\end{figure}

For our numerical calculation we approximately map the DMFT action onto a SIAM with a finite number $L_\text{bath}$  
of bath sites. Since we start from the atomic limit, there are no impurity-bath correlations 
in the initial state. Consequently the first bath 
vanishes, i.e., $\Lambda_-=0$, and we have $\Lambda=\Lambda_+$. Additionally, $\Lambda$ is 
spin symmetric ($\Lambda_\uparrow=\Lambda_\downarrow=\Lambda$) in the paramagnetic phase,
and particle-hole symmetric, such that we set up the SIAM symmetric as explained below \eqeqref{eq:decomp2},
with pairs of initially occupied and unoccupied bath orbitals:  
The  number of bath sites is $L_\text{bath}=2L$, where $L$ is the rank of the approximate representations of $i\Lambda^>$ 
and $-i\Lambda^<$, as introduced in Sec.~\ref{sec:approx_repres}. The initial ground state of the SIAM is sketched in 
Fig.~\ref{fig:initialstate} and contains an equal number of empty and doubly-occupied bath sites with energies 
$\epsilon_p=0$ and a singly-occupied impurity. 
In practice we average over two Green functions $G^\alpha$ and $G^\beta$, where the impurity of system 
$\alpha$ ($\beta$) is populated initially by a single up-spin (down-spin) electron, i.e., the lattice Green function is given by
\begin{align}
 G_{\sigma}(t,t')=\tfrac{1}{2}(G^\alpha_{0\sigma}(t,t')+G^\beta_{0\sigma}(t,t')).
\end{align}
Taking the average restores particle-hole symmetry, which is not given 
for $G^{\alpha}$ or $G^{\beta}$ alone.

The self-consistency condition for the Weiss field $\Lambda_\sigma$ is given by \eqeqref{bethe}. 
Because we will compare results from both eigenvector and Cholesky decomposition of $\Lambda$, we use
a DMFT iteration scheme with fixed maximum time $t_\text{max}$ [cf.~Fig. \ref{fig:dmft_schemes}a] (see also Fig.~\ref{fig:selfcon}). To this end, 
we initialize $\Lambda_n$ in iteration $n$ $=$ $1$ as
\begin{align}
	\label{eq:lambdainitial}
	\Lambda_1(t,t')
=v(t)g_0(t,t')v(t'),\quad t,t'\le t_\text{max},
\end{align}
where $g_0$ is a suitable initial Green function, e.g., the equilibrium Green function of the noninteracting Bethe lattice
which is known analytically.
After each decomposition of the Weiss field into hopping parameters $V_{0p}(t)$ we compute the real-time
impurity Green functions 
$G^s_{0\sigma}(t,t')=\Theta_C(t,t')G^{s,>}_{0\sigma}(t,t')+\Theta_C(t',t)G^{s,<}_{0\sigma}(t,t')$ with respect to the 
SIAMs $s=\alpha$ and $s=\beta$ by exact diagonalization (ED) techniques,
\begin{align}
 G^{s,>}_{0\sigma}(t,t')&=-i\bra{\psi_0^s}U(0,t)c_{0\sigma}U(t,t')c^\dagger_{0\sigma}U(t',0)\ket{\psi_0^s},&\nonumber\\
G^{s,<}_{0\sigma}(t,t')&=i\bra{\psi_0^s}U(0,t')c^\dagger_{0\sigma}U(t',t)c_{0\sigma}U(t,0)\ket{\psi_0^s},\nonumber\\
U(t,t')&=\text{T}_t\left\{\exP{-i\int_{t'}^{t}\mathrm{d}s\,H(s)}\right\},
\end{align}
where $\text{T}_t$ denotes the usual time-ordering operator. Due to the exponential growth of the Hilbert space 
(its dimension scales as $\binom{2L+1}{L}\binom{2L+1}{L+1}$) 
we use the Krylov method~\cite{lub:97} and a commutator-free exponential
time-propagation scheme~\cite{alv:11} to evolve the initial states $\ket{\psi_0^s}$ along the 
contour $C$. Also, we implemented fast updates for the time-dependent, sparse Hamiltonian matrices
and parallelize matrix-vector multiplications.

From the 
lattice Green function 
we can finally obtain the system's kinetic energy 
$\langle E_\text{kin}(t)\rangle=-i\sum_\sigma\int\limits_C\text{d}s\Lambda(t,s) G_\sigma(s,t')|^<_{t=t'}$ 
as well as the density $\langle n(t)\rangle=-i\sum_\sigma G^<_\sigma(t,t)$ which is a conserved quantity. Furthermore, the double occupation 
in the lattice is computed similarly to the Green function as the time-local impurity correlation function 
$\langle d (t)\rangle=\langle n_{0\uparrow}(t)n_{0\downarrow}(t)\rangle$ averaged over SIAMs $\alpha$ and $\beta$. The double occupation 
also gives access to the interaction energy, $\langle E_\text{int} (t)\rangle=U(\langle d(t)\rangle-\tfrac{1}{4})$.

\subsection{Comparison of eigenvector  and Cholesky approximation}
\begin{figure}[!t]
\includegraphics[width=\linewidth]{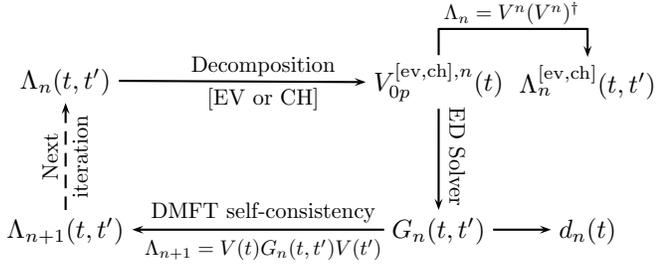}
\caption{DMFT iteration scheme for a fixed maximal time, i.e. $t,t'\le t_\text{max}$. The initial input Weiss field $\Lambda_{1}(t,t')$ is given by \eqref{eq:lambdainitial}. $d_n(t)$ denotes the double occupation after the $n$-th iteration.}
\label{fig:selfcon}
\end{figure}

In Sec.~\ref{sec:cmpcholev1} we emphasized that eigenvector and Cholesky approximation spread the error quite differently over the resulting approximate Weiss field. This has important consequences when we use them within the DMFT self-consistency cycle
explained in Fig.~\ref{fig:dmft_schemes}a,
and in detail in Fig.~\ref{fig:selfcon}.
\begin{figure}[!t]
	{\includegraphics[width=\linewidth]{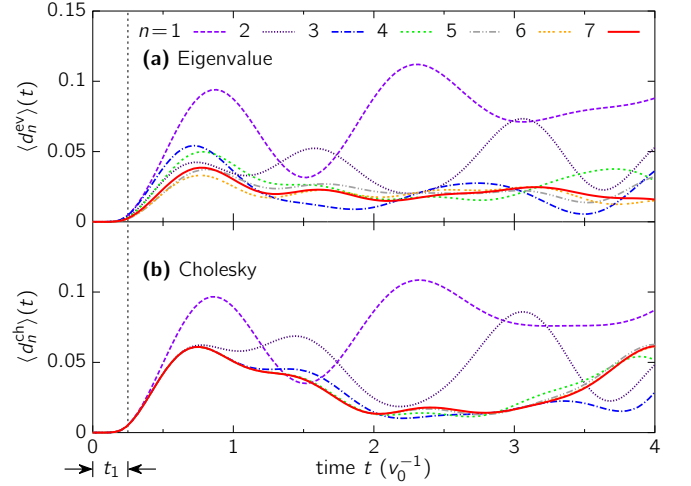}}\centering
	\caption{From every iteration we obtain a time evolution for the double occupation $d_n(t)$, represented by a single line in this plot. We used $t_\text{max}=4$, $L_\text{bath}=4$, $U=5$ and $v(t)$ as defined in \eqeqref{eq:ramp} with $t_1=0.25$ (vertical dotted line), cf. Fig. \ref{fig:double_occ} for its profile. $7$ iterations were performed. Top panel: Results obtained using the eigenvector approximation. Lower panel: Results obtained using the Cholesky approximation.}
\label{fig:causalityA}%
\end{figure}%
First differences between both approaches can be found by looking at intermediate results of the numerical calculation. In Fig.~\ref{fig:causalityA} we show results for the double occupation $d_n(t)$ after the $n$-th 
DMFT iteration.
For the calculation we fixed $t_\text{max}=4$ and used $L_\text{bath}=4$ bath sites for the SIAM. The curves in the top panel were calculated using the eigenvector approach. In each step one finds $\Delta^\text{ev}_n(t)\equiv d^\text{ev}_{n+1}(t)-d^\text{ev}_n(t)\neq 0$, except for very small times. This can be understood by recalling that the approximation $\Lambda^\text{ev}$ of the input Weiss field $\Lambda_n$ 
at DMFT iteration $n$
is calculated by discarding most of $\Lambda_n$'s eigenvectors. There are thus differences between $\Lambda_n(t,t')$ and 
its low rank approximation
for all times $t,t'$ and consequently $d_n(t)$ is affected as a whole.

Corresponding results for the Cholesky approach are plotted in the lower panel and reveal a causal nature. For each iteration one can find a time $t_{n}$ so that $\Delta^\text{ch}_n(t)\approx 0$ for $t<t_\text{n}$ (a converged calculation thus requires $t_n=t_\text{max}$). This can be understood as follows. Assume that the input Weiss field fulfills $\Lambda_{n}(t,t')=\Lambda_{n+1}(t,t')$ for $t,t'<t_n$. Due to the stepwise construction this leads to $\Lambda^\text{ch}_{n}(t,t')=\Lambda_{n+1}^\text{ch}(t,t')$ for $t,t'<t_n$ for the Cholesky approximation. The time evolution in a physical system is causal and therefore ensures $\Lambda_{n+1}(t,t')=\Lambda_{n+2}(t,t')$ for $t,t'<t_n$. This restricts all changes to $t,t'>t_n$.

\begin{figure}[!t]
{\includegraphics[width=\linewidth]{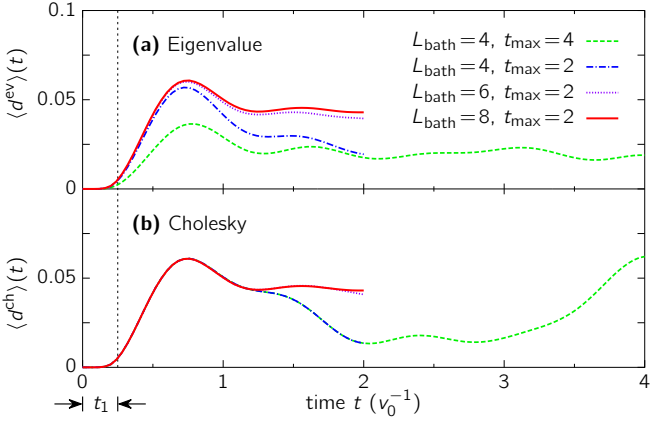}}\centering
\caption{Converged DMFT results obtained using eigenvector and Cholesky approximation. Calculations were performed for $U=5$ and $v(t)$ 
as defined in \eqeqref{eq:ramp}, with $t_1=0.25$ (vertical dotted line). 
Each line displays the final result for the double occupation $d(t)$, which depends on the maximal time $t_\text{max}$ 
and on the number of bath sites $L_\text{bath}$. Top panel: Eigenvector approximation. Lower panel: Cholesky approximation.}
\label{fig:causalityB}%
\end{figure}%
From these observations we conclude that the maximum time has to be chosen carefully if we work with the eigenvector approach. If $L_\text{bath}$ is chosen too small compared to the maximum time $t_\text{max}$, then we have to expect convergence against wrong results. Indeed we find such behavior as plotted in the upper panel of Fig. \ref{fig:causalityB}. The number of bath sites $L_\text{bath}=4$ turns out to be small if we consider $t_\text{max}=4$ and the resulting double occupation $d(t)$ 
(green dashed line)  
differs largely from the correct result 
(red solid line).

The stepwise construction of the Cholesky decomposition, on the other hand, ensures that one obtains correct results from the self-consistency cycle up 
to 
some maximum time, which
increases with the number of available bath sites $L_\text{bath}$ (Fig. \ref{fig:causalityB}, lower panel). In practice, the Cholesky approach 
appears to be preferable and was used for the calculation of all further results.

\subsection{Time evolution of energies and double occupation}
\label{sec:energies_and_double_occ}

\begin{figure}[!t]
\includegraphics[width=0.97\columnwidth]{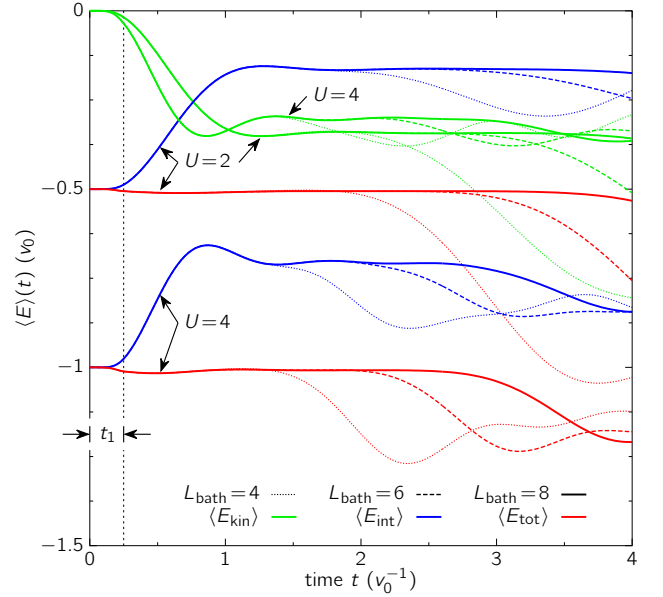}
\caption{Test of energy conservation. Numerical results for the time-dependent energies $\langle E_\text{kin}\rangle$ (green), $\langle E_\text{int}\rangle$ (blue) and $\langle E_\text{tot}\rangle$ (red) for $U=2$ (solid lines) and $U=4$ (dashed lines). The thick curves correspond to the SIAM representation with $L_\text{bath}=8$, the thinner curves to $L_\text{bath}=6$ and $4$ respectively. All data was obtained with the Cholesky decomposition. 
The vertical dotted line indicates the time $t_1$ at the end of the ramp, after which the Hamiltonian is
time-independent.
}
\label{fig:energy_cnsrv}
\end{figure}

Initially, in the atomic limit for times $t\leq0$, the electrons cannot hop between the lattice 
sites, there is no double occupation, and the system has total energy $-U/4$. 
Figure~\ref{fig:energy_cnsrv} 
shows the change of the kinetic, interaction and total energies
during and after the switch-on of the hopping for two selected values of the on-site interaction, $U=2$ and $U=4$. Also, we compare
results for different $L_\text{bath}$ in the SIAM representation.

The onset of the dynamics is characterized by a steady increase (decrease) 
of the absolute value of $\langle E_\text{kin}\rangle$ ($\langle E_\text{int}\rangle$). 
Note that the tiny energy transfer 
during the ramp does not imply that the system is in its ground state after the ramp.
The excitation energy in the system after the ramp (which would enter
an estimate of its effective temperature) is measured with respect to
the new ground state energy. For example, for a noninteracting system ($U$ $=$ $0^+$) we have $\est{n_{k\sigma}(t)}=\est{n_{k\sigma}(0)}=\frac{1}{2}$ and thus $E_\text{tot}(t)=0$, but
the ground state of the final Hamiltonian
is the Fermi sea with energy $\est{H(t_{1})}_\text{FS}=-\frac{8}{3\pi}<0$, so that the excitation energy is given by $\Delta E=\est{H(t_{1})}-\est{H(t_{1})}_\text{FS} =\frac{8}{3\pi}$ (for finite $U$ the value $-\frac{8}{3\pi}$ can serve as an upper bound for the ground-state energy of the final Hamiltonian $H(t_{1})$ because $\est{(n_\uparrow-\frac{1}{2})(n_\downarrow -\frac{1}{2})}_\text{FS}=0$.)

As an important 
check of the numerical results 
we note that the total energy is conserved after the switch-on is completed, 
including (for sufficiently large $L_\text{bath}>6$) the whole timescale on
which the energies saturate and approach a final value. On the other hand, we observe that the 
long-time dynamics for $t\gtrsim 3.5$ are not accurately described (in particular for larger $U$) 
because, instead of diverging results, we expect all energies to become constant. 
Clearly, in this regime the SIAM representation with few bath sites is inadequate and must be 
adjusted by increasing $L_\text{bath}$.

\begin{figure}[!t]
\includegraphics[width=0.97\columnwidth]{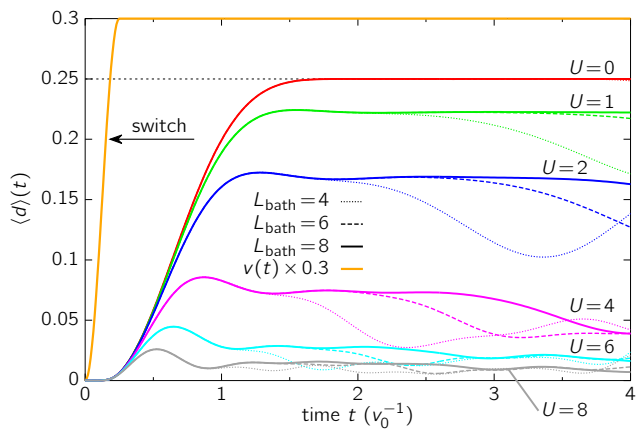}
\caption{Time dependence of the double occupation $d(t)$ in the Bethe lattice for the approximate SIAM representation with $L_\text{bath}=4$, $6$ and $8$ (using the Cholesky decomposition) and for different $U$ ranging from zero and small to large Coulomb interaction. The ramp of the hopping parameter (orange curve) is as in Fig.~\ref{fig:energy_cnsrv}.}
\label{fig:double_occ}
\end{figure}

In Fig.~\ref{fig:energy_cnsrv} the saturation of the time-dependent energies $\langle E_\text{kin}\rangle$ and $\langle E_\text{int}\rangle$ indicates the relaxation to a final steady state. An important
quantity of this state is its double occupation. 
Figure~\ref{fig:double_occ} 
shows how $d(t)$ builds up as function of time for different values of $U$. In the case of no interactions, $U=0$, we observe that the double occupation monotonically approaches a value of $d_\text{final}=1/4$ which is related to the ground state of the
paramagnetic phase in the presence of finite hopping $v_0$. Note that there is no energy input and $\langle E_\text{tot}\rangle(t)=\langle E_\text{kin}\rangle(t)=0$ for all times.
For increasing $U$ the final double occupation gets more and more suppressed. However, on the intermediate timescale, a pronounced switch-on behavior is formed resembling damped 
collapse and revival oscillations with approximate period $1/U$.

Regarding the calculations with different
numbers of bath sites, we conclude 
from Fig.~\ref{fig:double_occ} that for
$U\lesssim2$ a SIAM representation with
$L_\text{bath}=8$ is sufficient to correctly
resolve the dynamics up to $t$ $\lesssim$
$3$. On the other hand, the maximum
accessible time decreases with $U$, cf.\ the
curves for $U$ $=$ $4$, $6$, and $8$. We
attribute this to the initial atomic limit state,
for which the off-diagonal elements of the
Green functions $G^{<,>}(t,t')$ (and
hence eventually the hybridizations
$\Lambda^{<,>}(t,t')$) 
vary
on a timescale
proportional to $1/U$, i.e., more rapidly for
large $U$, thus requiring more bath sites to
capture hybridizations with higher rank
(cf.\ Sec.~\ref{sec:approxrepr}). {}From
a different perspective, the switches at larger
$U$ yield a smaller amount of excitation
energy, so that the time evolution involves
smaller energy differences and longer
timescales, the description of which
should be expected to require more
bath sites.

\subsection{Comparison to perturbation theory}

\begin{figure}[!t]
\includegraphics[width=0.97\columnwidth]{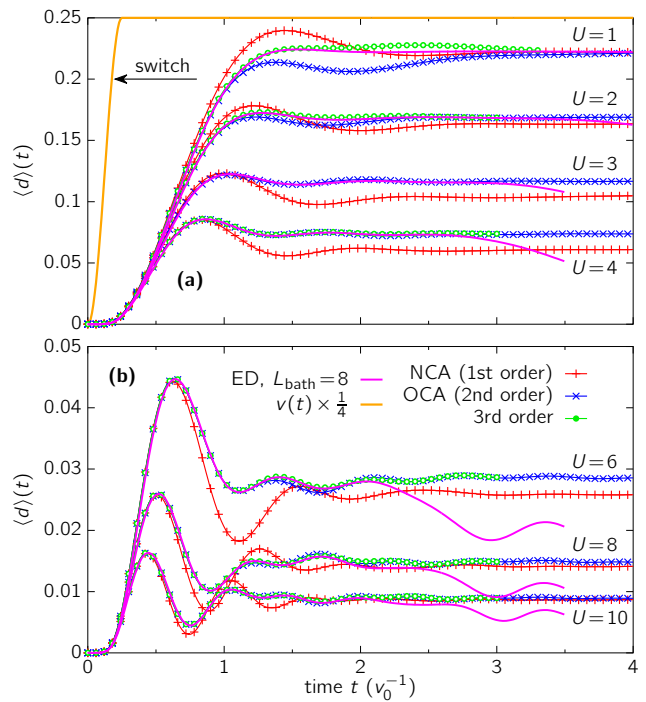}
\caption{Comparison of the SIAM-based results for $L_\text{bath}=8$ (violet curves) to the non-crossing approximation (NCA), the one-crossing approximation (OCA) and the corresponding third-order approximation of the hybridization expansion. While panel (a) covers small to moderate interactions, panel (b) addresses moderate to strong couplings. The ramp $v(t)$ (orange curve) is as in Figs.~\ref{fig:energy_cnsrv} and \ref{fig:double_occ}.}
\label{fig:nca_comparison}
\end{figure}

An essential advantage of using a Hamiltonian-based impurity solver in nonequilibrium DMFT is that 
results can be obtained independently of the strength of the Coulomb interaction. In the present case,
where the initial state is simple, the ED results can be considered as exact in the sense that they 
are converged with the number of bath orbitals for small
enough times. 
In Fig.~\ref{fig:nca_comparison}, we compare 
the
ED calculations of section~\ref{sec:energies_and_double_occ} to perturbative results based on 
a
hybridization expansion, which is most accurate when $U$ is much larger than the bandwidth.
The perturbative hybridization expansion 
uses a diagrammatic (``strong-coupling'') skeleton expansion of the impurity-bath hybridization around the 
atomic limit. The so-called non-crossing approximation (NCA) denotes the lowest-order variant of this conserving 
expansion, and  corrections to the NCA are 
obtained on the level of the one-crossing approximation (OCA). In Fig.~\ref{fig:nca_comparison}, in addition to the 
NCA and OCA we include calculations which sum the skeleton series up to third-order; for details see Ref.~\cite{Eckstein2010nca}.

The 
general observation in our case is that the NCA fails (as expected) for small $U$ but has also difficulties in describing 
the dynamics of $d(t)$ for interaction strengths as large as $U=10$. On the other hand, we find from Fig.~\ref{fig:nca_comparison}b, 
that low-order corrections to the NCA produce adequate results for $U>6$. The double occupation obtained from the OCA (and 
the expansion of third order which practically produces the same results) is in excellent agreement with the SIAM-based 
ED results for $L_\text{bath}$ $=$ $8$ up to  times for which the latter are reliable.

In the regime of moderate coupling (see the curves for $U=4$ and $6$ in Fig.~\ref{fig:nca_comparison}b), the OCA data start to deviate from the exact results 
such that the third-order approximation becomes indispensable to get convergent results. This trend intensifies if $U$ is further 
decreased. For small $U$ we find that even the third-order approximation cannot reproduce the ED curves.
Note that the numerical effort for evaluating the third order diagrams is already quite substantial,
as it involves multiple integrations over the contour $C$ and thus scales like $\mathcal{O}(N^5)$ with the 
number of time-discretization steps \cite{Eckstein2010nca}. Weak-coupling methods such as CTQMC, on the other hand, 
cannot be used easily because they cannot describe the atomic limit initial state.

\section{Summary and outlook}
\label{sec:summary}

In this paper we have addressed the problem of representing the
nonequilibrium DMFT action by a time-dependent Hamiltonian, i.e., a
single-impurity Anderson model (SIAM). The solution of this ``mapping
problem'' makes it possible to adopt powerful wave-function based
numerical time-propagation algorithms, such as Krylov-space methods or
DMRG, as impurity solvers for nonequilibrium DMFT.

To solve the mapping problem, we determined a SIAM with time-dependent
impurity-bath hopping parameters $V_{0p}(t)$, which has the same
hybridization function $\Lambda(t,t')$ as the DMFT action. In
equilibrium, $\Lambda$ is uniquely determined by a positive-definite
spectral function to which the parameters of the Hamiltonian can be
fitted. By contrast, for a nonequilibrium Green function with two time
arguments $t$ and $t'$ on the Keldysh contour and no
time-translational invariance, even the representability by a SIAM is
not obvious. In this paper we have used a Lehmann representation to
prove that the representability is a rather general property of
nonequilibrium Green functions, and we have presented an explicit
procedure to construct the parameters of the SIAM.

Using this scheme, we found that the number of bath degrees of freedom
that are needed for an accurate representation of a given $\Lambda$
typically increases with the maximum evolution time, beginning with
the representation of the initial equilibrium state. The bath orbitals
belong to two different classes (denoted as first and second bath in
this paper), which differ both in their physical meaning and in the
way in which the respective SIAM parameters are determined. The first
bath is coupled to the impurity already in the initial state at time
$t=0$ and it is determined by fitting a generalized time- and
frequency-dependent spectral function related to the mixed
imaginary-time/real-time sector of the hybridization function, i.e.,
these parameters account for correlations between the initial state at
$t\le0$, and times $t>0$. In a general nonequilibrium situation, we
find that the first bath must gradually decouple from the impurity in
order to correctly represent the fading memory of the initial state in
the DMFT action. The second bath, on the other hand, is gradually
coupled to the impurity at times $t>0$ and describes correlations
which are build up at later times.  Numerically, we determine it using
a matrix decomposition (eigenvector or Cholesky decomposition) in real
time.

Both the Cholesky and the eigenvector decomposition allow us to find
an approximate bath representations by minimizing the difference
between the exact $\Lambda$ and the representation based on a finite
number of bath orbitals only. The Cholesky decomposition has the
advantage that the optimal SIAM at a given time can be determined
uniquely from the solution at previous times (and for the initial
equilibrium state), such that the resulting impurity solver can be
embedded into a step-wise time propagation scheme.  Furthermore, this
fact ensures that even with a small number of bath sites we can obtain
converged results for not too large times. In a first numerical test
of the approach, we have studied the time evolution after a ramp of
the hopping in the Hubbard model, starting from the atomic limit. This
setup, which is in principle easy to realize with cold atoms, is hard
to solve both with weak-coupling perturbation theory or weak-coupling
CTQMC (which cannot describe the initial state), and with
strong-coupling perturbation theory (which is no longer accurate when
$U$ becomes smaller than the hopping).

There are several open questions and future research directions.
First of all, it would be interesting to analyze how the number of
bath orbitals that are needed for a given accuracy scales with the
maximum evolution time. We have not investigated this question
systematically, because the maximum number of bath sites that can be
dealt within exact diagonalization techniques is rather limited
anyway. For conceptual reasons, however, and for the use of different
solvers, this question is certainly relevant.

Moreover, based on our proof of the representability of the DMFT
action by a SIAM, various other procedures of finding the actual SIAM
appear worthwhile to study. For example, instead of minimizing the
difference between the impurity hybridization function and a
hybridization function in the DMFT action, one may minimize the
difference between the DMFT Green function $G_\text{lat}$ (computed
from the lattice Dyson equation with the impurity self-energy) and the
impurity Green function $G_\text{SIAM}$. The difference between
$G_\text{lat}$ and $G_\text{SIAM}$ might actually be a better error
measure than the corresponding difference between the hybridization
functions $\Lambda$ and $\Lambda_\text{SIAM}$: $G_\text{SIAM}(t,t')$
is a superposition of exponentially many oscillating components whose
dephasing leads to a rapid decay of the function at large
time differences $t-t'$. It is thus typically already closer to the
Green function of an infinite system ($G_\text{lat}$) than
$\Lambda_\text{SIAM}$ is to $\Lambda$. However, the numerical
implementation will be more complicated because the approximation error
is computed between two quantities that are both determined
numerically.

Finally, any method that relies on an ad-hoc truncation of the bath
can potentially violate conservation laws of energy or particle
number. There are two possible ways to cure this problem (apart from
increasing the number of bath sites): First of all, a systematic
perturbation theory around the truncated impurity model can be
formulated in the language of dual-fermions~\cite{Jung2012}.  The
approach can be made conserving and it can also potentially alleviate
other finite-bath size effects, but the evaluation of the
corresponding diagrams requires considerable numerical effort. Another
very interesting direction is the nonequilibrium generalization of
self-energy functional theory~\cite{potthoff2003,potthoff2003c}, where
the parameters of the impurity model are not determined ad hoc, but
from the stationary point of the dynamical Luttinger-Ward variational
principle. Although the numerical implementation of the approach is
slightly more challenging, it is a worthwhile endeavor because spin and
particle number conservation would be satisfied in such
an approach with a suitable choice of the variational space~\cite{Hoffmann2013}.

As a next step, it would certainly be worthwhile to combine the
mapping procedure with efficient numerical schemes such as DMRG, which
might be able to reach timescales that are not accessible with any of
the currently available solvers.  The use of Hamiltonian-based solvers
for nonequilibrium DMFT has only just begun, and based on the
theoretical foundations presented here, several interesting topics can
be expected to be explored soon.

\acknowledgments

C.G. and M.K. were supported in part by Transregio~80 of the Deutsche
Forschungsgemeinschaft.

\appendix
\section{Role of the bath geometry}
\label{ap:role}
In the following discussion we consider a bath with arbitrary geometry
and show that one can always find a SIAM with the same number of sites
and a simple star geometry which has the same effective action.
For simplicity we suppress the spin index. The associated bath Hamiltonian is given by 
\begin{equation}
	H_\text{bath}=\sum_{p,p'>0}(V_{pp'}(t)-\mu\delta_{pp'})\ads{p}\aas{p'}.
\end{equation}
$V_{pp'}(t)$ is a hermitian matrix and can be diagonalized at every time $t$. For $t=0$ we call the necessary unitary transform $\mat{O}$, so that
\begin{equation}
	\mat{V}(0)=\mat{O} \mat{D} \mat{O}^\dagger\quad\text{with}\quad D_{pp'}=\delta_{pp'}d_p.
\end{equation}
Corresponding to the bath Hamiltonian is the noninteracting one-particle Green function ($S_\text{bath}\equiv-i\int_C \dt H_\text{bath}(t)$; cf.~\eqref{gdef} for the definition of the expectation value)
\begin{multline}
	g_{pp'}(t,t')
\equiv-i\est{\aas{p}(t)\ads{p'}(t')}_{S_\text{bath}}\\
=i\left[\mat{U}(t,0)\left(f[\mat{V}(0)-\mu]-\Theta_C(t,t')\right)\mat{U}^\dagger(t',0)\right]_{pp'},
	\label{eq:MatFerm}
\end{multline}
where (for $t>t'$)
\begin{equation}
	\mat{U}(t,t')=\torder{\Exp{-i\int\limits^t_{t'} \left(\mat{V}(t_1)-\mu\right)\dt_1}}.
	\label{eq:picard}
\end{equation}
The expression $f[\mat{V}(0)-\mu]$ refers to the matrix Fermi distribution, i.e.,
\begin{align}
	&f[\mat{V}(0)-\mu]=f[\mat{O}(\mat{D}(0)-\mu)\mat{O}^\dagger]=\mat{O}f[\mat{D}(0)-\mu]\mat{O}^\dagger,\nonumber\\
	&f[\mat{D}(0)-\mu]_{pp'}=\delta_{pp'}f(d_p-\mu).
\end{align}
$f(\epsilon)$ is the Fermi function. For the following consideration we assume that $V_{pp'}(t)$ are chosen to fulfill
\begin{equation}
	\Lambda(t,t')=\sum_{p,p'>0}V_{0p}(t)g_{pp'}(t,t')V_{0p'}^*(t'),
	\label{eq:identlambda}
\end{equation}
i.e., the bath Hamiltonian describes a valid mapping. To show the equivalence of the arbitrary geometry with a star structure we search for a diagonal bath Hamiltonian with constant eigenenergies that also reproduces the Weiss field $\Lambda(t,t')$.

We define the quantity
\begin{equation}
	v_{p}(t)\equiv e^{i (d_p-\mu) t}\sum_{\tilde{p},p'>0}V_{0\tilde{p}}(t)U_{\tilde{p}p'}(t,0)O_{p'p},
	\label{eq:notunique}
\end{equation}
which ensures $v_{p}(-i\tau)=v_{p}(0)$ for every $\tau$. This can be seen by taking a look at the propagator $\mat{U}(-i\tau,0)$, which can be written as
\begin{align}
	\mat{U}(-i\tau,0)&=\Exp{-i\int\limits^{-i\tau}_{0} \left(\mat{V}(t_1)-\mu\right)\dt_1}\nonumber\\
	&=\mat{O}\,\Exp{-(\mat{D}-\mu)\tau}\mat{O}^\dagger.
\end{align}
Note that $\mat{V}(-i\tau)=\mat{V}(0)$ follows from $\mat{V}$'s definition as hopping matrix of the bath Hamiltonian. Inserting this into~\eqref{eq:notunique} cancels the unitary transformation $\mat{O}$ on the right and one finds 
\begin{align}
	v_{p}(-i\tau)&= e^{ (d_p-\mu) \tau}\sum_{\tilde{p},p'>0}V_{0\tilde{p}}(0)O_{\tilde{p}p'}\Exp{-(\mat{D}-\mu)\tau}_{p'p}\nonumber\\
	&=\sum_{\tilde{p}>0}V_{0\tilde{p}}(0)O_{\tilde{p}p}=v_{p}(0).
	\label{eq:validbath}
\end{align}
It is therefore possible to use $v_p(t)$ as the hopping parameter for a new geometry. But first we notice that our definition leads to the following diagonal form for $\Lambda(t,t')$, i.e.,
\begin{equation}
	\Lambda(t,t')=\sum_{p>0}v_{p}(t)h_{p}(t,t')v_{p}(t')^*.
	\label{eq:diagident}
\end{equation}
$h_{p}(t,t')$ is the Green function of the time-independent diagonal bath ($\mat{V}(t)\rightarrow\mat{D}(0)$) describing a noninteracting equilibrium situation
\begin{equation}
	\label{eq:nonintgreen}
	h_{p}(t,t')= i(f(d_{p}-\mu)-\Theta_C(t,t'))\text{e}^{-i (d_{p}-\mu)(t-t')}.
\end{equation}
We would have found the exact same expression for $\Lambda(t,t')$ if we had started from the following bath and hybridization matrices
\begin{align}
	H^\text{star}_\text{bath}&=\sum_{p>0}(d_{p}-\mu)\ads{p}\aas{p},
\nonumber\\
        H^\text{star}_\text{hyb}&=\sum_{p>0}(v_{p}(t)\ads{0}\aas{p}+\text{h.c.}).
\end{align}
This results suggests that there is no advantage in choosing a different, more complicated bath geometry. The Green function $g_{pp'}(t,t')$ can be stated analytically only for the star structure, where it is diagonal. For an arbitrary geometry one has to deal with the $\mat{U}(t,t')$ which can only be calculated numerically from $V_{pp'}(t)$ and vice versa. We also emphasize that the star-structured bath Hamiltonian involves the same amount of lattice sites. This results from the construction of $v_p(t)$ by a unitary transform. Therefore, if we are approximate $\Lambda(t,t')$ using a finite number of sites, e.g., in a numerical calculation, there is no advantage in choosing a complicated geometry.

\section{Cavity method for the effective action}
\label{sec:genf}

Here we obtain the action~\eqref{eq:impurityaction}, i.e. we evaluate the definition \eqref{eq:Stildedef} of $\tilde{S}$
\begin{align}
	\exp(\tilde{S})&=\sum_{n=0}^\infty\frac{1}{n!}\est{(\Delta S)^n}_{S^{(0)}}.
	\label{eq:bathtrace}
\end{align}
Keeping in mind that the contour ordering for operators acting on $\mathcal{F}_0$ still has to be performed, cf. \eqref{eq:part2}, we will treat the operators $\cds{0\sigma},\ccs{0\sigma}$ as (anticommuting) constants when evaluating the expectation value. From the definition of $\Delta S$ we conclude that only terms with an equal number of $\cds{i\sigma},\ccs{j\sigma}$ with $i,j\neq 0$ are nonzero and thus only terms which are of an even power of $\Delta S$ contribute. For easier notation we will suppress the spin index for creation (annihilation) operators acting on $\mathcal{F}_\text{rest}$ and the hopping in the following (it can be reinserted using $\ccs{i_n}\rightarrow\ccs{i_n\sigma_n},\cds{j_m}\rightarrow\cds{j_m\sigma'_m}, t_{ij}\rightarrow t_{ij}^\sigma$). For the operators at site 0 we drop the site index, i.e., $\ccs{0\sigma}\rightarrow \ccs{\sigma}$. We define the $n$-particle contour-ordered Green function as
\begin{equation}
	G^{(0)}_{i_1\dots i_n,j_1,\dots,j_n}(t_1,\dots,t'_n)\equiv(-i)^n\est{\ccs{i_1}(t_1)\dots\cds{j_n}(t'_n)}_{S^{(0)}}.
\end{equation}
The contour ordering is again contained in the definition of the expectation value, cf.~\eqref{eq:S(0)def}.

To evaluate~(\ref{eq:bathtrace}) let us consider which terms are generated by $(\Delta S)^{(2n)}$. From the definition of $\Delta S$ we can see that only product terms where $n$ operators $\ccs{i}$ are multiplied with $n$ operators $\cds{j}$ are not equal to zero. Each term can be constructed by choosing $n$ creation operators $\cds{j}$ out of the $2n$ possibilities. This fixes the annihilation operators one has to choose exactly, so that ${(2n)!}/{n!^2}$ $n$-particle Green function contribute. When we order those operators to match the definition of an $n$-particle Green function, we also reorder the operators at site $0$, so that the sequence of time variables is the same. For instance, a term in lowest order takes the form
\begin{equation}
 \cds{j}(t)\ccs{\sigma}(t)\cds{\sigma'}(t')\ccs{i}(t')=\ccs{i}(t')\cds{j}(t)\cds{\sigma'}(t')\ccs{\sigma}(t).
\end{equation}
Terms of higher order can be thought of as a product of $n$ such terms and one readily realizes that the total sign is not affected by the reordering. The definition of the Green function consumes a factor $(-i)^n$ so the same factor is left (from $(\Delta S)^{2n}$ we got a $(-i)^{2n}$). In the end we find
\begin{multline}
	\label{eq:defZ}
	\exp(\tilde{S})
        =\sum_{n=0}^\infty\frac{1}{(2n)!}\est{(\Delta S)^{2n}}_{S^{(0)}}\\
	=\sum_{n=0}^\infty\int\limits_C\dt_1\dots\int\limits_C\dt_n'\sum_{i_1,\dots,j_n}(-i)^n\frac{t_{0i_1}(t_1)\dots  t_{j_n0}(t_n')}{n!n!}
	\\
        \times G^{(0)}_{i_1,\dots,j_n}(t_1,\dots,t'_n)\cds{\sigma_1}(t_1)\dots\ccs{\sigma'_n}(t'_n).
\end{multline}

We proceed to reexponentiate the right-hand side of~\eqref{eq:defZ} using connected (with respect to the interaction in $H^{(0)}(t)$, cf.~\eqref{eq:hamexpansion}) contour-ordered Green functions. For a easier notation we define
\begin{multline}
  G_n\equiv\int\limits_C\dt_1\dots\int\limits_C\dt_n'\sum_{i_1,\dots,j_n}t_{0i_1}\dots  t_{j_n0}\\
  \times G^{(0)}_{i_1,\dots,j_n}(t_1,\dots,t'_n)\cds{\sigma_1}(t_1)\dots\ccs{\sigma'_n}(t'_n).
\end{multline}
and in an analogous way $G_n^\text{c}$, i.e., $G^{(0)}\rightarrow G^{(0),c}$, so that
\begin{equation}
	\exp(\tilde{S})=\sum_{n=0}^\infty\frac{(-i)^n G_n}{n!n!}.
	\label{eq:genfun}
\end{equation}
An $n$-particle Green function can be written entirely in terms of connected $m$-particle (with $m\le n$) Green functions. We consider first one single contributing term of $s$ connected $m_k$-particle Green functions (of course, $\sum_{k=1}^s m_k=n$ has to hold for such a summand). The $n$-particle Green function consists of $n$ creation and $n$ annihilation operators. This implies that there are ${n \choose m_1}^2$ possibilities to construct a $m_1$-particle connected Green function out of it. For the next there are only $n-m_1$ creation (annihilation) operators left, so that there are ${{n-m_1} \choose m_2}^2$ possibilities to create the $m_2$-particle connected Green function. Using this scheme we find the following contribution to $G_n$:
\begin{multline}
	\!\!\!\!\!\!{{n} \choose m_1}^2{{n-m_1} \choose m_2}^2\dots{{n-m_1-\dots-m_{s-1}} \choose m_s}^2\prod_{k=1}^s G_{m_k}^\text{c}\\
	=n!n!\prod_{k=1}^s \frac{G_{m_k}^\text{c}}{m_k!m_k!}.
\end{multline}
The definition of $G_n$ and $G_n^\text{c}$ ensures that we do not have to think about sign issues here. An operator reordering might be necessary to match the definition of a connected Green function but we also have to reorder the operators at site $0$ in the exact same way, so that the total sign does not change.

As the next step we generate all the other summands. We start from terms that are a product of $s=1$ connected Green functions to terms which consist of $s=n$ connected Green functions. This way we find
\begin{equation}
	G_n=n!n!\sum_{s=1}^n\frac{1}{s!}\sum_{m_1=1}^\infty\dots\sum_{m_s=1}^\infty\prod_{k=1}^s \frac{G_{m_k}^\text{c}}{m_k!m_k!}\delta_{\sum_{k=1}^s m_k,n},
\end{equation}
where the Kronecker delta $\delta_{\sum_k m_k,n}$ selects all the contributions which fulfill the condition \makebox{$\sum_{k=1}^s m_k=n$}. The factor $\frac{1}{s!}$ takes care of the fact that the order of the factors does not matter (e.g., $G_1^c G_2^c=G_2^c G_1^c$). We can now plug this result into~\eqref{eq:genfun}
\begin{multline}
	\exp(\tilde{S})=1+\sum_{n=1}^\infty\sum_{s=1}^n\frac{1}{s!}
\\\times
       \sum_{m_1=1}^\infty\dots\sum_{m_s=1}^\infty\prod_{k=1}^s \frac{(-i)^{m_k} G_{m_k}^\text{c}}{m_k!m_k!}\delta_{\sum_{k=1}^s m_k,n}.
\end{multline}
To get the final result we have to regroup the summands in this expression by the number of factors they consist of. This is easily done by extending the $s$-summation to infinity, which does not introduce extra terms because of the Kronecker delta. The $n$-summation is now readily calculated and one finds the desired result
\begin{align}
	\exp(\tilde{S})&=1+\sum_{s=1}^\infty\frac{1}{s!}\sum_{m_1=1}^\infty\dots\sum_{m_s=1}^\infty\prod_{k=1}^s \frac{(-i)^{m_k} G_{m_k}^\text{c}}{m_k!m_k!}\\
	&=\sum_{s=0}^\infty\frac{1}{s!}\left(\sum_{m=1}^\infty\frac{(-i)^m G_{m}^\text{c}}{m!m!}\right)^s
\nonumber\\&
        =\exP{\sum_{m=1}^\infty\frac{(-i)^m G_{m}^\text{c}}{m!m!}},\nonumber
	\label{eq:wlnz}
\end{align}
completing the derivation of~\eqref{eq:impurityaction} and~\eqref{eq:deflam}.

\section{The limit of infinite lattice dimension and locality of the self-energy}
\label{ap:selfcondarblatt}
In this appendix we apply the limit of infinite lattice-dimension to simplify the effective local action \eqref{eq:impurityaction}. This will lead us to the DMFT action $S_\text{loc}$, cf. \eqref{dmft g}. Based on the cavity formalism we will then show that the lattice self-energy is local and establish the DMFT self-consistency condition.

\subsection{DMFT action for an infinite-dimensional lattice}
As in the equilibrium case~\cite{Georges96}, the hybridization functions~\eqref{eq:deflam} simplify drastically in the limit of infinite
lattice dimension, $d$ $\to$ $\infty$.
Applying the quantum scaling~\cite{Metzner1989}
$t_{ij}\propto {d}^{-\mathcal{Z}_{ij}}$ ($\mathcal{Z}_{ij}$ is the number of sites connected by hopping $t_{ij}$) it can be shown by counting
powers of $d$ that only first-order terms (i.e., one-particle Green
functions) contribute to the effective action. We now show this for the special case of
nearest-neighbor hopping on a hypercubic lattice, using standard arguments~\cite{Metzner1989,Georges96}.

We consider the contributions to~\eqref{eq:deflam} from Green
functions of $n$th order, for which the lattice summations yield a total factor $d^{2n}$.
We have $t_{0i}\propto d^{-1/2}$ for the  nearest neighbors of site $0$ and thus the product of $2n$ such quantities contributes a factor $d^{-n}$. The Green functions in~\eqref{eq:deflam} only contain connected diagrams. Because all the sites it connects are nearest neighbors of $0$ the shortest path between them is of length $2$ (using the metric $||\vec{R}||=\sum_i \abs{R_i}$). To connect $2n$ sites we need at least $2n-1$ such paths and thus the largest terms are of order $(\sqrt{d})^{2(2n-1)}=d^{2n-1}$ if all sites are different. Using these results on~(\ref{eq:deflam}) leads to
\begin{multline}
	\Lambda_{\sigma_1\dots\sigma_n'}(t_1,\dots,t_n')\propto\\
	\underbrace{\sum_{i_1,\dots,j_n}}_{\propto\,d^{2n}}\underbrace{t_{0i_1}(t_1)\dots t_{j_n0}(t_n')}_{\propto (\sqrt{d})^{-2n}}\underbrace{G^{(0),\text{c}}_{(i_1 \sigma_1),\dots,(j_n \sigma_n')}(t_1,\dots,t'_n)}_{\propto (\sqrt{d})^{-2(2n-1)}}
\\
\propto \frac{1}{d^{n-1}}.
\end{multline}
If only $2n-m$ sites are different, the order reduces to $2(2n - m - 1)$. However, this constraint also reduces the factor given by the summation to $d^{2n-m}$. In total, we always find $\Lambda_{\sigma_1\dots\sigma_n'}\propto d^{n-1}$, which proves that only first-order terms ($n$ $=$ $1$) contribute. We further have $\Lambda_{\sigma\sigma'}(t,t)=\delta_{\sigma\sigma'}\Lambda_{\sigma}(t,t')$ since the Hubbard Hamiltonian does not involve spin flips. In the limit $d$ $\to$ $\infty$ the effective action thus reduces 
to~\eqref{eq:dmftaction} with hybridization~\eqref{eq:cavity}. The partition function is given by $Z_\text{loc}=\traceo{\tcorder{\exP{S_\text{loc}}}}$ and describes an impurity that is coupled to the Weiss field $\Lambda_{\sigma}(t,t)$. This Weiss field represents the influence of all other lattice sites.

\subsection{Local self-energy and self-consistency condition for an arbitrary lattice}
Here we use the cavity formalism to establish that the self-energy is local for an  arbitrary lattice in the limit of infinite dimensions and derive the corresponding self-consistency condition. We begin by replacing
\begin{equation}
	t_{i0}(t)\ccs{}(t)\rightarrow\eta_{i}(t),\quad t_{0i}(t)\cds{}(t)\rightarrow\eta_{i}^\dagger(t),
\end{equation}
in~\eqref{eq:actionS}, with $\{\eta_i(t),\ccs{j}\}=\{\eta_i(t),\cds{j}\}=0$ for $i,j\neq 0$. We further assume that, within a contour-ordered expression, all $\eta_i(t)$ anticommute pairwise with each other and with $\ccs{},\cds{}$. They thus operate on the impurity site ($i=0$; we recall the definition $\ccs{}\equiv \ccs{0}$). One easily verifies that all steps of the derivation in Appendix~\ref{sec:genf} remain valid (even the limit $d$ $\to$ $\infty$) and we conclude
\begin{multline}
	F[\bullet,\eta,\eta^\dagger]\equiv\tcorder{\bullet\,\,\exP{S_\text{loc}[\eta,\eta^\dagger]}}\\
        =\frac{1}{Z^{(0)}}\tcorder{\,\bullet\,\,\traceb{\exP{S[\eta,\eta^\dagger]}}}\equiv F_\text{lat}[\bullet,\eta,\eta^\dagger],
\end{multline}
where ``$\bullet$'' represents an arbitrary product of time-dependent operators acting on site 0. We recall that $S[\eta,\eta^\dagger]=S_0+S^{(0)}+\Delta S[\eta,\eta^\dagger]$ and
\begin{align}
	\Delta S[\eta,\eta^\dagger]&=-i\int\limits_C\dt\left[\sum_{i\neq 0}\left(\cds{i}(t)\eta_i(t)+\eta^\dagger_i(t) \ccs{i}(t)\right)\right],\nonumber\\
	S_\text{loc}[\eta,\eta^\dagger]&=-i\int\limits_C\dt\int\limits_C\dt'\sum_{ij\neq 0}G_{ij}^{(0)}(t,t')\eta^\dagger_i(t)\eta_j(t')+S_0.\nonumber
\end{align}

As for Grassmann variables, a functional derivative with the common rules (chain rule, product rule etc.) and ($\xi\in\{\eta,\eta^\dagger\}$)
\begin{equation}
	\label{eq:funcderrel}
	\acomm{\frac{\delta}{\delta \xi_i(t)},c_j}=\acomm{\frac{\delta}{\delta \xi_i(t)},c^\dagger_j}=0,\,\frac{\delta \xi_i(t)}{\delta \xi_j(t')}=\delta_C(t,t')\delta_{ij},
\end{equation}
with the contour delta function $\delta_C(t,t')\equiv \partial_t\Theta_C(t,t')$, can now be defined as follows. Note that it is not allowed to choose $\delta\eta_j(t)$ proportional to the unit matrix, as this would violate the anticommutation requirement. However, it is possible to define an ``anticommuting unit matrix'' by introducing an additional dummy site with corresponding creation (annihilation) operators $f^\dagger$ ($f$). For a functional $F[\xi]$, with $\xi\in\{\eta,\eta^\dagger\}$, we then set 
\begin{align}
\frac{\delta F[\xi]}{\delta \xi_j(t)}&\equiv\text{tr}_f\left\{f^\dagger\lim_{\epsilon\rightarrow0}\frac{F[\xi+\epsilon\delta_C(\cdot,t)\delta_{\cdot,j}f]-F[\xi]}{\epsilon}\right\},
\nonumber\\
\delta\xi_j(t)&\equiv\epsilon\delta_C(\cdot,t)\delta_{\cdot,j}f.
\end{align}
Note that no dummy time variable for $f$ is needed since it anticommutes with every other involved creation (annihilation) operator except $f^\dagger$. Here $\text{tr}_f$ traces over the subspace of the dummy site. The differential operator $\frac{\delta}{\delta \xi_j(t)}$ indeed anticommutes with every other fermionic creation (annihilation) operator [cf.~\eqref{eq:funcderrel}].

Since the functionals $F[\bullet,\eta,\eta^\dagger]$ and $F_\text{lat}[\bullet,\eta,\eta^\dagger]$ coincide everywhere, any functional derivative of them is equal. Note also that both functionals contain only operators that act on the impurity site. It is now straightforward to verify that (for $i,j\neq0$)
\begin{align}
	G_{ij}(t,t')&=\frac{-i}{Z_\text{loc}}\left.\traceo{\frac{\delta^2 F_\text{lat}[1,\eta,\eta^\dagger]}{\delta\eta^\dagger_i(t)\delta\eta_j(t')}}\right|_{\substack{\eta_k(t)=t_{k0}(t)c(t) \\ \eta^\dagger_l(t)=t_{0l}(t)c^\dagger(t)}}\nonumber\\
	&=\frac{-i}{Z_\text{loc}}\left.\traceo{\frac{\delta^2 F[1,\eta,\eta^\dagger]}{\delta\eta^\dagger_i(t)\delta\eta_j(t')}}\right|_{\substack{\eta_k(t)=t_{k0}(t)c(t) \\ \eta^\dagger_l(t)=t_{0l}(t)c^\dagger(t)}}\nonumber\\
	&=G^{(0)}_{ij}(t,t')+\int\limits_C\!\!\dt_1\int\limits_C\!\!\dt_2\sum_{kl}G_{ik}^{(0)}(t,t_1)t_{k0}(t_1)
       \nonumber\\&\phantom{=}\;\;
        \times G(t_1,t_2)t_{0l}(t_2)G_{lj}^{(0)}(t_2,t').
	\label{eq:startgij}
\end{align}
In this context $G_{ij}(t,t')=-i\est{\ccs{i}(t)\cds{j}(t')}_S$ is the lattice Green function [cf.~(\ref{eq:actionS})] and $G(t,t')=G_{00}(t,t')$ the Green function at the impurity. We arrive at the equation
\begin{multline}
	\sum_{ij}t_{0i}(t)G_{ij}(t,t')t_{j0}(t')=\\
	\Lambda(t,t')+\int\limits_C\dt_1\int\limits_C\dt_2\;\Lambda(t,t_1)G(t_1,t_2)\Lambda(t_2,t').
\end{multline}
To close the self-consistency for a given $\Lambda(t,t')$, we need an expression for the lattice Green function $G_{ij}(t,t')$. To derive it we consider
\begin{align}
	G_{0j}(t,t')&=\frac{1}{Z_\text{loc}}\left.\traceo{\frac{\delta F[c(t),\eta,\eta^\dagger]}{\delta\eta_j(t')}}\right|_{\substack{\eta_k(t)=t_{k0}(t)c(t) \\ \eta^\dagger_l(t)=t_{0l}(t)\cds{}(t)}}\nonumber\\
	&=\frac{1}{Z_\text{loc}}\left.\traceo{\frac{\delta F_\text{lat}[\ccs{}(t),\eta,\eta^\dagger]}{\delta\eta_j(t')}}\right|_{\substack{\eta_k(t)=t_{k0}(t)c(t) \\ \eta^\dagger_l(t)=t_{0l}(t)c^\dagger(t)}}.\nonumber
\end{align}
The evaluation yields ($j$ $\neq$ $0$)
\begin{align}
	\label{eq:secondrel}
        G_{0j}(t,t')=\int\limits_C\dt_1\sum_i G(t,t_1)t_{0i}(t_1)G^{(0)}_{ij}(t_1,t').
\end{align}
A similar conjugated equation can be derived for $G_{i0}(t,t')$, which gives after summation 
\begin{align}
	\sum_i t_{0i}(t) G_{i0}(t,t')=\int\limits_C \dt_1 \Lambda(t,t_1)G(t_1,t').
	\label{eq:necfordyson1}
\end{align}
Insertion of~\eqref{eq:secondrel} into~\eqref{eq:startgij} leads to
\begin{multline}
	\label{eq:noneqanalog}
	G^{(0)}_{ij}(t,t')=G_{ij}(t,t')\\
       -\int\limits_C\dt_1\int\limits_C\dt_2 G_{i0}(t,t_1)G^{-1}(t_1,t_2)G_{0j}(t_2,t'),
\end{multline}
where the (matrix) inverse (with respect to time arguments) $G^{-1}(t,t')$ is defined by the impurity Dyson equation
\begin{align}
	\int\limits_C\dt_1G^{-1}(t,t_1)G(t_1,t')=\delta_C(t,t').
	\label{eq:impdyson}
\end{align}
Note that~\eqref{eq:noneqanalog} is the analogue of the equilibrium relation $G^{(0)}_{ij}(i\omega_n)=G_{ij}(i\omega_n)-{G_{i0}(i\omega_n)G_{0j}(i\omega_n)}/{G(i\omega_n)}$, i.e., (36) in~\cite{Georges96}. 

The inverse $G^{-1}(t,t')$ of the impurity Green function $G(t,t')$ is connected to the impurity self-energy $\Sigma(t,t')$ by 
\begin{align}
	G^{-1}(t,t')&=(i\partial_t + \mu)\delta_C(t,t') - \Lambda(t,t')-\Sigma(t,t'),
\end{align}
and allows one to rewrite~(\ref{eq:secondrel}), with $j\neq0$, as
\begin{equation}
	\int\limits_C \dt_1 G^{-1}(t,t_1) G_{0j}(t_1,t')=\sum_{i} t_{0i}(t)G^{(0)}_{ij}(t,t').
	\label{eq:needinv}
\end{equation}
By multiplying~\eqref{eq:noneqanalog} with $t_{0i}(t)$ and summing over $i$, we find
\begin{multline}
	\sum_i t_{0i}(t) G_{ij}(t,t')\\
=\int\limits_C \dt_1 [G^{-1}(t,t_1) + \Lambda(t,t_1)]G_{0j}(t_1,t')
	\label{eq:necfordyson2}
\end{multline}
for $j\neq0$, where we used~\eqref{eq:necfordyson1} and~\eqref{eq:needinv}. Now the inverse lattice Green function $G^{-1}_\text{lat}(t,t')$ is defined by the lattice Dyson equation
\begin{align}
	\sum_k\int\limits_C\dt_1 (G^{-1}_\text{lat})_{ik}(t,t_1)G_{kj}(t_1,t')=\delta_{ij}\delta_C(t,t'),
\end{align}
and thus is connected to the lattice self-energy $\Sigma_\text{lat}(t,t')$ via 
\begin{multline}
	(G^{-1}_\text{lat})_{ij}(t,t')\equiv[\delta_{ij}(i\partial_t+\mu)-t_{ij}(t)]\delta_C(t,t') 
\\
       -(\Sigma_\text{lat})_{ij}(t,t').
\end{multline}
Note that here the inverse with respect to both the time arguments and the lattice indices appears.  

Now we can show that $(\Sigma_\text{lat})_{ij}(t,t')=\delta_{ij}\Sigma_\text{loc}(t,t')$, i.e., the lattice self-energy is local and given by the self-energy of the impurity model. We have 
\begin{align}
	\delta_{0j}\delta_C(t,t')&=\sum_k\int\limits_C\dt_1 (G^{-1}_\text{lat})_{0k}(t,t_1)G_{k j}(t_1,t')\\
	&=(i\partial_t+\mu)G_{0j}(t,t')-\sum_k t_{0k}(t)G_{k j}(t,t')\nonumber\\
	&\phantom{=}\,\,-\int\limits_C \dt_1 \sum_k (\Sigma_\text{lat})_{0k}(t,t_1)G_{k j}(t_1,t')\nonumber
\end{align}
and by using~\eqref{eq:necfordyson1} and~\eqref{eq:necfordyson2} we identify (for arbitrary $j$)
\begin{multline}
	\int\limits_C\dt_1 \Sigma(t,t_1)G_{0j}(t_1,t')
\\
        =\int\limits_C\dt_1\sum_k (\Sigma_\text{lat})_{0k}(t,t_1)G_{k j}(t_1,t').
	\label{eq:neconesite}
\end{multline}
Since we consider a translationally invariant system, the same result has to hold for an arbitrary site $i$, i.e.,
\begin{multline}
	\int\limits_C\dt_1 \Sigma(t,t_1)G_{ij}(t_1,t')
\\
        =\int\limits_C\dt_1\sum_k (\Sigma_\text{lat})_{ik}(t,t_1)G_{k j}(t_1,t').
\end{multline}
Applying the inverse of the lattice Green function from the r.h.s, we find the desired result
\begin{equation}
	(\Sigma_\text{lat})_{ij}(t,t')=\delta_{ij}\Sigma(t,t').
\end{equation}
This result can be used to calculate $G_{ij}(t,t')$ from the lattice Dyson equation and thus closes the self-consistency condition.
We note that the same argument
 is valid in case of a system that is not translationally invariant. In this case one has to calculate a local action $(S_\text{loc})_i$ for each site $i$. This allows one to derive expression (\ref{eq:neconesite}) for each site separately, and one finds the more general result
\begin{equation}
	(\Sigma_\text{lat})_{ij}(t,t')=\delta_{ij}\Sigma_i(t,t').
\end{equation}

For completeness we express the self-consistency condition by means of the impurity Dyson equation,
\begin{multline}
  \label{eq:vornedysonimp}
  \int\limits_C\dt_1\;[(i\partial_t + \mu)\delta_C(t,t_1) - \Lambda(t,t_1)
\\
-\Sigma(t,t_1)]G(t_1,t')=\delta_C(t,t').
\end{multline}
the lattice Dyson equation,
\begin{multline}
\label{eq:vornedysonlatt}
\int\limits_C\dt_1 
	\sum_k       \Big[
[\delta_{ik}(i\partial_t+\mu)-t_{ik}(t)]\delta_C(t,t_1)
\\
-\Sigma(t,t_1)\Big]G_{k j}(t_1,t')=\delta_{ij}\delta_C(t,t'),
\end{multline}
with $G_{00}$ $=$ $G$. 
For given hybridization $\Lambda$ and $G$ (with $G$ calculated from the SIAM, see next subsection), one can obtain $\Sigma$ from~\eqref{eq:vornedysonimp}, and use~\eqref{eq:vornedysonlatt} to obtain a new $G$, and thus a new $\Lambda$ from~\eqref{eq:vornedysonimp}.
In practice, one Fourier transforms~\eqref{eq:vornedysonlatt} to momentum space. Eqs.~\eqref{eq:vornedysonimp}-\eqref{eq:vornedysonlatt} are known, e.g., from Ref.~\onlinecite{Freericks2006,Eckstein11}, where their numerical evaluation is discussed.

\section{Analytical properties of nonequilibrium Green functions}
\label{sec:anaprop}
In this section we summarize the analytical properties of the Matsubara and mixed components~\eqref{eq:mixr} and~\eqref{eq:mixl}  of 
the contour-ordered Green functions (\ref{gdef}),
following Ref.~\onlinecite{Eckstein2010}. 
For simplicity of notation, we assume that the time evolution is determined by a time-dependent Hamiltonian, i.e., the 
action is $S=-i\int_C \text{d}s H(s)$. Note that this includes the general case in which the time-nonlocal part of the action is 
representable by a SIAM. We will use the convention that operators with a hat are in the Heisenberg picture, 
\begin{equation}
	\label{ap:defHeisenberg}
	\cc{i}(t)=U(0,t)\cc{i}U(t,0), \quad\cc{i}\equiv \cc{i}(0)=\ccs{i}.
\end{equation}
where $U(t,t')$ is the propagator associated with the system. Since we consider thermal initial states, their analytical properties are 
very similar to those of equilibrium Green functions (see, e.g.,~\cite{wal:81}). 

We start by recalling the properties of the Matsubara Green function~\eqref{gdef mat}.
It can be Fourier transformed 
using  fermionic Matsubara frequencies \makebox{$\omega_n=\frac{(2n+1)\pi}{\beta}$},
\begin{align}
	&G_{ij}^\text{M}(\tau-\tau')
\label{eq:generic} =\frac{1}{\beta}\sum_{n}e^{-i\omega_n(\tau-\tau')}g^\text{M}_{ij}(i\omega_n),
\\
	&\text{with}\quad g^\text{M}_{ij}(i\omega_n)=\int\limits_0^\beta \text{d}\tau\, e^{i\omega_n\tau}G^\text{M}_{ij}(\tau).
\end{align}
The Fourier components $g^{\text{M}}_{ij}(i\omega_n)$ can be analytically continued into the upper or lower 
complex frequency plane.
By the requirement $g^\text{M}_{ij}(z)\stackrel{z\to\infty}{\propto} \frac{1}{|z|}$ it is ensured that this continuation 
is equal to the Laplace transform of the retarded (for $\text{Im}(z)>0$) or advanced (for $\text{Im}(z)<0$) Green function~\cite{wal:81}. 
By expanding $G^\text{M}(\tau)$ using the eigenbasis $\rwf{n}$ with $E_n$ as the corresponding eigenenergy of the initial 
Hamiltonian $H(0)$ one obtains the Lehmann representation for this function,
\begin{equation}
	\label{eq:anacon}
	g^\text{M}_{ij}(z)=
	\sum_{m,n}
	\frac{e^{-\beta E_n}+e^{-\beta E_m}}{Z}\,
	\frac{\lwf{n}\cd{i}\rwf{m}\lwf{m}\cc{j}\rwf{n}}{z-(E_m-E_n)}.
\end{equation}
The function $g^\text{M}(z)$ has a branch cut at the real axis which is purely imaginary and 
related to
the spectral function $A_{ij}(\omega)$, 
\begin{align}
	\label{eq:spectral}
	A_{ij}(\omega)&\equiv \frac{i}{2\pi}[g^\text{M}_{ij}(\omega+i0)-g^\text{M}_{ij}(\omega-i0)]\\
					 &=\sum_{mn}\frac{e^{-\beta E_n}+e^{-\beta E_m}}{Z}
					 \nonumber\\&\phantom{=\sum}\;\;\times
					 \lwf{n}\cd{i}\rwf{m}\lwf{m}\cc{j}\rwf{n}\delta(\omega-(E_m-E_n))\nonumber.
\end{align}
In turn, $G^M$ is uniquely determined by its spectrum,
\begin{align}
g_{ij}^\text{M}(z) 
&= 
\int\limits_{-\infty}^\infty \text{d}\omega \,\frac{A_{ij}(\omega)}{z-\omega}
\\
G_{ij}^\text{M}(\tau) 
&=
\int\limits_{-\infty}^\infty \text{d}\omega\, A_{ij}(\omega) (f(\omega)-\Theta(\tau)) e^{-\omega\tau},
\end{align}
where $\Theta(\tau)$ denotes the Heaviside step function.

The generalization for the mixed components is straightforward.
We introduce a partial Fourier series,
\begin{align}
	\label{eq:fourierconv}
	G^\mixl(i\omega_n,t)&=\int\limits_0^\beta\text{d}\tau\, G^\mixl(\tau,t)e^{i\omega_n\tau},\\
	G^\mixr(t,i\omega_n)&=\int\limits_0^\beta\text{d}\tau\, G^\mixr(t,\tau)e^{-i\omega_n\tau},\\
	G^\mixl(\tau,t)&=\frac{1}{\beta}\sum_n G^\mixl(i\omega_n,t)e^{-i\omega_n\tau},\\
	G^\mixr(t,\tau)&=\frac{1}{\beta}\sum_n G^\mixr(t,i\omega_n)e^{i\omega_n\tau}.
\end{align}
Both can be analytically continued in the lower and upper complex plane. We restrict the discussion to 
$G^\mixl_{ij}(z,t')$:
Using the same scheme as for~\eqref{eq:anacon} we obtain
\begin{align}
	G^\mixl_{ij}(z,t')&=
	i
	\sum_{m,n}
	\frac{e^{-\beta E_n}+e^{-\beta E_m}}{Z}\,
	\frac{\lwf{n}\cd{i}\rwf{m}\lwf{m}\cc{j}(t')\rwf{n}}{z-(E_m-E_n)}.
\end{align}
A time-dependent generalization to the spectral function is given by the branch cut of this function 
along the real axis,
\begin{align}
	\label{eq:tspectral}
	A^\mixl_{ij}(\omega,t)
	&\equiv 
	\frac{1}{2\pi}\left[G^\mixl_{ij}(\omega+i0,t')-G^\mixl_{ij}(\omega-i0,t)\right]\\
					 &=\frac{1}{Z}\sum_{mn}\left(\exP{-\beta E_n}+\exP{-\beta E_m}\right)
       \nonumber\\&\phantom{\equiv}\;\;
       \times\lwf{n}\cd{i}\rwf{m}\lwf{m}\cc{j}(t)\rwf{n}\delta(\omega-(E_m-E_n))\nonumber.
\end{align}
To ensure $A^\mixl_{ij}(\omega,0)=A_{ij}(\omega)$ we have omitted the factor $i$ in this definition.
The corresponding relations for $G^\mixr_{ij}(t,z)$ and $A^\mixr_{ij}(t,\omega)$ can be determined from
$G^\mixr_{ij}(t,z)$ $=$ $(G^\mixl_{ji}(z^*,t))^*$ and its consequence $A^\mixr_{ij}(t,\omega)$ $=$ $(A^\mixl_{ji}(\omega,t))^*$.
Although $A^\mixl_{ij}(\omega,t)$ is not a real and positive function, it uniquely determines the mixed components
\begin{align}
	\label{eq:lambdacon2}
	G^{\mixr}(t,\tau')&=i\int\limits_{-\infty}^\infty d\omega \,A^\mixr(t,\omega)f(\omega)\exP{\omega\tau'},\\
	G^{\mixl}(\tau,t')&=i\int\limits_{-\infty}^\infty d\omega \,A^\mixl(\omega,t')(f(\omega)-1)\exP{-\omega\tau}.\nonumber
\end{align}

For a time-independent Hamiltonian the generalization of the spectral function is trivial. The additional time dependence just yields a phase factor
\begin{equation}
	\label{eq:trivtime}
	A^\mixl_{ij}(\omega,t')=\exP{i\omega t'}A_{ij}(\omega).
\end{equation}

For the investigation of the mapping problem the analytical properties of the Weiss field $\Lambda_{\sigma}(t,t')$ are of importance. 
From the cavity representation (\ref{eq:cavity}) one can see that the hybridization function has the same analytical properties
as any Green function.

\section{Positive definiteness of $-i\Lambda_+^<$ and $i\Lambda_+^>$}
\label{ap:pdlambdap}
According to Sec.~\ref{sec:zerotempbath} the construction of the second bath $V^+_{0p}(t)$ comes down to a matrix decomposition which requires $-i\Lambda_+^<$ and $i\Lambda_+^>$ to be positive definite. We will show that the assumption that $\Lambda$ is representable by a SIAM ensures the positive definiteness of $-i\Lambda_+^<$ and $i\Lambda_+^>$.

For simplicity we suppress the spin index. We start from~\eqref{eq:actionreq} and define the density of states
\begin{equation}
	\rho(\epsilon)=\frac{1}{L}\sum_p\delta(\epsilon-\epsilon_p),\text{ with } L=\int\limits_{-\infty}^\infty\text{d}\epsilon\sum_p \delta(\epsilon-\epsilon_p).
\end{equation}
To a given energy $\epsilon$ we define the hybridization function $v_k(\epsilon,t)\equiv\sqrt{L}\,V_{0p_k(\epsilon)}(t)$, where the sequence $p_k(\epsilon)$ runs through all $p$ with $\epsilon_p=\epsilon$. This way~\eqref{eq:actionreq} can be written as
\begin{equation}
	\label{ap:pdintv}
	\Lambda(t,t')=\sum_k\int\limits_{-\infty}^\infty\text{d}\epsilon \rho(\epsilon)v_k(\epsilon,t)g(\epsilon-\mu,t,t')v^*_k(\epsilon,t').
\end{equation}
We interprete $v_k(\epsilon,t)$ as the $k$-th component of a vector $\vec{v}(\epsilon,t)$, so that $(\vec{v}(\epsilon,t))_k=v_k(\epsilon,t)$.~\eqref{ap:pdintv} thus takes the form
\begin{equation}
	\Lambda(t,t')=\int\limits_{-\infty}^\infty\text{d}\epsilon\rho(\epsilon) \vec{v}^\dagger(\epsilon,t')\vec{v}(\epsilon,t)g(\epsilon-\mu,t,t').
\end{equation}
Following the steps~\eqref{eq:followme1}-\eqref{eq:mixed} of Sec.~\ref{sec:memofinit} we find
\begin{equation}
	C^\mixr(t,\epsilon)=\exP{-i\epsilon t}\rho(\epsilon)\vec{v}^\dagger(\epsilon+\mu,0) \vec{v}(\epsilon+\mu,t)
\end{equation}
and conclude that $\Lambda_{-}(t,t')$ is of the form
\begin{align}
	\Lambda_{-}(t,t')&=i\int\limits_{-\infty}^\infty\text{d}\epsilon\rho(\epsilon)\, \vec{v}^\dagger(\epsilon,t')P_{-} \vec{v}(\epsilon,t)g(\epsilon-\mu,t,t'),\nonumber\\
	P_{-}&=\frac{\vec{v}(\epsilon,0)\,\vec{v}^\dagger(\epsilon,0)}{\abs{\vec{v}(\epsilon,0)}^2}.
\end{align}
The operator $P_{-}$ projects onto the subspace spanned by $\vec{v}(\epsilon,0)$. We define a corresponding operator $P_+$ that projects onto the orthogonal complement space
\begin{equation}
	P_+=1-P_{-}.
\end{equation}
This way we find
\begin{align}
	\Lambda_{+}(t,t')&=i\int\limits_{-\infty}^\infty\text{d}\epsilon\rho(\epsilon)\, \vec{v}^\dagger(\epsilon,t')P_{+} \vec{v}(\epsilon,t)g(\epsilon-\mu,t,t')
\end{align}
It is now easy to verify that $-i\Lambda^<_+$ and $i\Lambda^>_+$ are indeed positive definite.

\end{document}